\documentclass[aps,prx,superscriptaddress,floatfix,onecolumn,nofootinbib,notitlepage]{revtex4-1}

\usepackage{graphicx}
\usepackage{dcolumn}
\usepackage{bm}
\usepackage{amsmath, amsthm}
\usepackage{rotating}
\usepackage{amsmath,amsthm,amssymb,mathtools}
\usepackage{algorithm}
\usepackage[noend]{algpseudocode}
\usepackage{verbatim}
\usepackage{float}
\usepackage{subcaption}
\usepackage{outline}
\usepackage{csquotes}
\usepackage{tikz}
\usetikzlibrary{shapes.geometric, arrows}
\usepackage{braket}
\usepackage{colonequals}
\usepackage{textcomp}
\usepackage[inline]{enumitem}
\usepackage[colorlinks]{hyperref}
\usepackage[all]{hypcap}

\newlist{inlinelist}{enumerate*}{1}
\setlist*[inlinelist,1]{%
  label=(\roman*),
}
\makeatletter
\def\BState{\State\hskip-\ALG@thistlm}
\makeatother

\usepackage{color}
\usepackage{xcolor}

\newcommand{\eq}[1]{Eq.~\hyperref[eq:#1]{(\ref*{eq:#1})}}
\renewcommand{\sec}[1]{\hyperref[sec:#1]{Section~\ref*{sec:#1}}}
\DeclareRobustCommand{\app}[1]{\hyperref[app:#1]{Appendix~\ref*{app:#1}}}
\newcommand{\tab}[1]{\hyperref[tab:#1]{Table~\ref*{tab:#1}}}
\newcommand{\fig}[1]{\hyperref[fig:#1]{Figure~\ref*{fig:#1}}}
\newcommand{\figa}[2]{\hyperref[fig:#1]{Figure~\ref*{fig:#1}#2}}
\newcommand{\figx}[2]{\hyperref[fig:#1]{Figure~\ref*{fig:#1}(#2)}}
\newcommand{\thm}[1]{\hyperref[thm:#1]{Theorem~\ref*{thm:#1}}}
\newcommand{\lem}[1]{\hyperref[lem:#1]{Lemma~\ref*{lem:#1}}}
\newcommand{\cor}[1]{\hyperref[cor:#1]{Corollary~\ref*{cor:#1}}}
\newcommand{\defn}[1]{\hyperref[def:#1]{Definition~\ref*{def:#1}}}
\newcommand{\alg}[1]{\hyperref[alg:#1]{Algorithm~\ref*{alg:#1}}}

\DeclareMathOperator*{\argmin}{\arg\!\min}

\def\avg#1{\mathinner{\langle{#1}\rangle}}
\def\bra#1{\mathinner{\langle{#1}|}}
\def\ket#1{\mathinner{|{#1}\rangle}}

\input{Qcircuit}

\begin{document}
\title{Application of fermionic marginal constraints to hybrid quantum algorithms}

\author{Nicholas C. Rubin}
\email{nick@rigetti.com}
\affiliation{Rigetti Computing, Berkeley, CA 94710}
\author{Ryan Babbush}
\email{ryanbabbush@gmail.com}
\affiliation{Google Inc., Venice, CA 90291}
\author{Jarrod McClean}
\email{jarrod.mcc@gmail.com}
\affiliation{Google Inc., Venice, CA 90291}

\begin{abstract}
Many quantum algorithms, including recently proposed hybrid classical/quantum algorithms, make use of restricted tomography of the quantum state that measures the reduced density matrices, or marginals, of the full state.  The most straightforward approach to this algorithmic step estimates each component of the marginal independently without making use of the algebraic and geometric structure of the marginals.  Within the field of quantum chemistry, this structure is termed the fermionic $n$-representability conditions, and is supported by a vast amount of literature on both theoretical and practical results related to their approximations.  In this work, we introduce these conditions in the language of quantum computation, and utilize them to develop several techniques to accelerate and improve practical applications for quantum chemistry on quantum computers. As a general result, we demonstrate how these marginals concentrate to diagonal quantities when measured on random quantum states.  We also show that one can use fermionic $n$-representability conditions to reduce the total number of measurements required by more than an order of magnitude for medium sized systems in chemistry. As a practical demonstration, we simulate an efficient restoration of the physicality of energy curves for the dilation of a four qubit diatomic hydrogen system in the presence of three distinct one qubit error channels, providing evidence these techniques are useful for pre-fault tolerant quantum chemistry experiments.
\end{abstract}

\maketitle
\section*{Introduction}

Quantum computers are expected to provide an exponential speedup in the solution of the electronic structure problem ~\cite{Lloyd1996,Abrams1997,Abrams1999,Ortiz2001,Somma2002,Aspuru-Guzik2005} in some cases. This could potentially revolutionize research in chemistry and material science by allowing a new mechanism for designing new materials, drugs, and catalysts. Accordingly, there is now a significant body of literature dedicated to developing new algorithms \cite{Kassal2008,Ward2009,Whitfield2010,Jones2012,Toloui2013,Hastings2015,BabbushSparse2,BabbushSparse1,BabbushAQChem,Kivlichan2016,Sugisaki2016,BabbushLow,Motzoi2017,Kivlichan2017}, tighter bounds and better implementation strategies \cite{Veis2010,Veis2014,McClean2014,Wecker2014,Poulin2014,BabbushTrotter,Reiher2017,Berry2018}, more desirable Hamiltonian representations \cite{Seeley2012,Whitfield2013b,Tranter2015,Moll2015,Whitfield2016,Barkoutsos2017,Havlicek2017,Bravyi2017,Zhu2017,Setia2017,Steudtner2017}, and experimental
demonstrations~\cite{Lanyon2010,Li2011,Wang2014,Santagati2016} for this problem.

The rapid development of quantum hardware in recent years has spurred interest in the development of practical algorithms that do not require fault-tolerant quantum computers.  It has been conjectured that the leading candidates for first demonstrations of practical algorithms on a pre-threshold device are so-called quantum/classical hybrid algorithms, such as the variational quantum eigensolver (VQE) for chemistry \cite{Peruzzo2013,Yung2013,McClean2015,Shen2015,Wecker2015a,Sawaya2016,OMalley2016,Kandala2017,Romero2017} and the quantum approximate optimization algorithm (QAOA) for optimization problems \cite{Farhi2014}.  Recent experimental results \cite{OMalley2016} have shown that the quantum-classical feedback does provide robustness against noise in the device, and simple extensions are possible that allow one to further damp the noise through additional measurements \cite{McClean2016,Siddiqi2017}.

A common step in almost all near-term implementations of these algorithms is the determination of an operator's expected value through a form of partial tomography of the quantum state.  The statistical criteria relating the number of measurements from the distribution and the accuracy of the expected value are relatively straightfoward to determine~\cite{McClean2014, McClean2015, Romero2017}.  When studying chemical systems, the general benchmark for accuracy is $\langle H \rangle  \pm  \epsilon$ Hartree where $\epsilon = 1.6 \times 10^{-3}$ is known as chemical accuracy and $\pm \epsilon$ expresses a standard confidence interval. This stringent absolute accuracy is required for matching experimentally determined thermochemical properties such as heats of formation or ionization potentials~\cite{RevModPhys.71.1267}.  Naturally, this accuracy requirement on the estimator for the Hamiltonian may require a large number of independent preparations of a state and measurements using current strategies.

Some routes to reduce the number of experiments by collecting Hamiltonian terms into commuting groups and dropping Hamiltonian terms with small coefficients have already been proposed~\cite{raeisi2012quantum,McClean2014,McClean2015,Kandala2017}. Wecker \textit{et al.}~\cite{Wecker2015a}, estimated that for simulating the energy of ferrodoxin using this approach would require $10^{19}$ total measurements.  Though this estimate was made with pessimistic assumptions, the sheer number of measurements motivates one to pursue techniques to accelerate the operator averaging step of VQE and other hybrid algorithms.

The strategy we adopt in order to ameliorate the burden of excessive measurements is to leverage known structure in the marginals of the fermionic density matrices which have not yet been taken advantage of within the field of quantum computation. The set of necessary conditions on marginals of density matrices are known as $n$-representability conditions and were originally developed to use the reduced density matrices of fermionic  and bosonic systems as the main computational variable instead of the wavefunction~\cite{RevModPhys.32.170, coleman2000reduced, coleman1978reduced, cioslowski2000many}. While in this work we focus on fermionic systems with at most pair-wise interactions and thus use constraints on the $2$-marginal $^{2}D$, these ideas are more generally applicable to problems with local Hamiltonian objectives, such as many instances of QAOA.  Measuring the marginals of a state $\rho$, specifically the fermionic two-particle reduced-density-matrix ($2$-RDM), provides a powerful extension to VQE.  These quantities are useful for connecting VQE to other quantum chemical techniques such as multiconfigurational self-consistent field (MCSCF), embedding procedures such as density matrix embedding theory, and augmenting the accuracy of electronic structure methods with perturbation theory, which is thought to be required to apply these methods to complex systems larger than one may fit in the quantum computer alone.

In this work we explore the utility of measuring marginals instead of directly measuring the Hamiltonian, and how the $n$-representability constraints on the marginals provide \textit{i}) a program for reducing the norm of the Hamiltonian directly leading to fewer measurements and \textit{ii}) using the set of necessary, but not sufficient constraints, on the fermionic marginals, known as $p$-positivity constraints, to propose two computational procedures for projecting measured marginals into the set of allowed marginals.  The variance reduction that is developed is examined for linear hydrogen chains and generally shows an order of magnitude reduction in the required number of measurements. 

We also introduce a series of polynomial time post-processing techniques for certifying and projecting measured marginals.  This computational procedure is similar to maximum likelihood tomography except on a reduced density matrix space, thus making it efficient.  Similar techniques have been applied to corrupted process tomography~\cite{foley2012measurement} and reduced state tomography on the RDMs of the system~\cite{cramer2011efficient}.  We compare four projection techniques that balance enforcing $n$-representability with computational efficiency.   The first two techniques are based on positive projection of the $2$-RDM matrix--which should be positive semidefinite--but does not include any constraints beyond positivity of the $2$-RDM and fixing the trace.  The third and fourth techniques add approximate $n$-representability constraints implemented through a more expensive iterative procedure and semidefinite program.

The paper is structured as follows: Sections \ref{sec:marginals} and \ref{sec:nrep} describe fermionic marginals and the $n$-representability conditions used in this work, Sections \ref{sec:concentration} and \ref{sec:rdm_pertubation} review the utility of marginals as they pertain to perturbation theory and a concentration bound for $2$-RDMs that suggests a structured ansatz is required when parameterizing the space for hybrid quantum classical algorithms, Section \ref{sec:op_av_bound} discusses the optimal bound on the number of samples required to measure the expected value of a Hamiltonian and variance reduction from equality $n$-representability constraints, Section \ref{sec:nrep_proj} discusses the $2$-RDM projection procedures and how they can reduce the number of measurements required to measure to fixed accuracy $\epsilon$ and also restore physicality after state corruption by common error channels. We close with an outlook as to how $n$-representability techniques can further improve hybrid algorithms.

\section{Marginals of the density operator}\label{sec:marginals}
Here we introduce our terminology and set notation with respect to reduced density matrices, or marginals of the full density, in both the qubit and fermionic setting.  Generally speaking, a marginal of a multivariable probability distribution is the partial trace, or integration, of a subset of the variables leaving a distribution on a smaller set of variables.  Given a general quantum state $\rho$ on $n$ qubits 
\begin{align}
\rho = \sum_{i}w_{i}\vert \psi_{i} \rangle \langle \psi_{i} \vert
\end{align}
where $\vert \psi_{i}\rangle$ are pure states of $n$ qubits and $\sum_{i}w_{i} = 1$, the set of $p$-qubit reduced-density matrices, or $p$-marginals, of the state are determined by integrating out $q$-qubits (such that $n - q = p$) of the joint distribution as
\begin{align}\label{eq:qubitdensity}
^{p}\rho_{m_{1}, ..., m_{p}} = \mathrm{Tr}_{n_{1}, n_{2}, ..., n_{q}}[\rho]
\end{align}
resulting in $\binom{n}{p}$ different marginals each of dimension $2^{p} \times 2^{p}$.  The coefficients $n_{1}, ..., n_{q}$ on the trace operator indicate which qubits are integrated out of $\rho$ and coefficients $m_{1}, ..., m_{p}$ label the subsystem marginal.  The result of marginalization is a distribution on the state space of $p$-qubits.  We will interchangeably refer to these objects as marginals and reduced-density matrices (RDMs) alluding to the fact that like the von Neumann density matrix the marginals can be expressed as a sum 
\begin{align}
^{p}\rho_{m_{1}, ..., m_{p}} = \sum_{j}w_{j}\vert \phi_{j} \rangle \langle \phi_{j} \vert
\end{align}
where $\vert \phi_j \rangle$ is a pure state on the subsystem of qubits that has not been traced out.

It is well known that this $\binom{n}{p}$ set of marginals is sufficient for calculating the expected value of a $p$-local Hamiltonian or other observable~\cite{liu2006consistency, bravyi2003requirements}. A $p$-local Hamiltonian is one where any term in the Hamiltonian involves no more than $p$-qubits interacting. Naturally, such a description of a quantum system is an attractive polynomial size representation.  However, in order to use the set of marginals as computational objects (e.g. minimize energy with respect to them) rather than simply measuring them, the RDMs must satisfy certain constraints to ensure physicality.  These constraints are termed ``consistency'' and are the requirement that all marginals satisfy Eq.~\eqref{eq:qubitdensity} for the same initial state $\rho$.  Despite the considerable structure on $^{p}\rho_{m_{1}, ..., m_{p}}$, confirming a set of marginals are consistent was demonstrated to be  QMA-complete~\cite{liu2006consistency}.  Naturally, one may ask if there is some alternative form of consistency, or approximation, that makes working with marginal distributions a computationally attractive approach.

Analogous to the case of qubits, the requirement for calculating expected values of $k$-local operators describing interactions of indistinguishable particles, such as fermions or bosons, is that only the $k$-marginal is needed.  Note that as a point of clarification, while $p$-local qubit operators refer to operators acting on at most $p$ qubits, $k$-local fermionic operators refer to interactions that derive from $k$-body interactions, and generically act on $2k$ fermionic modes.  This does not complicate the methods to be discussed, but is a common point of confusion when working between chemistry and quantum computation.  
As an example, in chemical systems the energy is a functional of the $1$- and $2$-local fermionic operators and the $2$-marginal of the system~\cite{szabo1989modern, mazziotti2002variational, coleman1980reduced, coleman2000reduced}. 
Consider a Fock space constructed with a single particle basis associated with a Hilbert space $\mathcal{H}$ of size $m$.  We can represent a general state on this space
\begin{align}\label{eq:fermion_state}
\vert \psi \rangle = \sum_{i_{1}, ..., i_{m} = 0}^{1}c_{i_{1},...,i_{m}}(a_{m}^{\dagger})^{i_{m}}(a_{m-1}^{\dagger})^{i_{m-1}}...(a_{1}^{\dagger})^{i_{1}} \vert \mathrm{vac} \rangle
\end{align}
where each fermionic creation operator $\{a_{m}^{\dagger}\}$ is associated with a single particle basis state $\vert \phi_{m} \rangle$ and $\vert \mathrm{vac}\rangle$ is the vacuum.
We may consider a state $\psi$ with fixed particle number $n$ enforced by restricting the coefficients $i_{m}$ in Eq~\eqref{eq:fermion_state} to satisfy $\sum_{m}i_{m} = n$.  As noted in Reference~\cite{verstraete2005mapping}, once an ordering of fermions is selected the Fock space states can be mapped to a Hilbert space of $m$-distinguishable spin-$\frac{1}{2}$ particles.  Marginals of the fermionic $n$-particle density matrix
\begin{align}
^{n}D = \vert \psi \rangle \langle \psi \vert = {}^{n}D_{j_{1}, ..., j_{n}}^{i_{1},...,i_{n}} \vert i_{n} \dotsm i_{1} \rangle \langle j_{n} \dotsm j_{1} \vert 
\end{align}
involve integrating out particles by a trace operation.  For example, the $2$-RDM ${}^{2}D$ is obtained from ${}^{n}D$ by integrating out particles $3$ to $n$.
\begin{align}
{}^{2}D = \mathrm{Tr}_{3...n}[{}^{n}D].
\end{align}
By convention, in the quantum chemistry community the normalization constant for $^{p}D$ is typically scaled to $\binom{n}{p}$ when $i_{1} < i_{2} < ... < i_{p}$ or $n!/(n - p)!$ when $i_{k}$ is allowed to be range over all values in $[1, m]$~\cite{coleman2000reduced, nakata2001variational, cioslowski2000many}. In this work we choose the later of the normalizations for ease of computation.  

From the study of $n$-representability theory, a number of efficiently implementable and necessary constraints on the one- and two-particle marginals (${}^{1}D$ and ${}^{2}D$) are known. Specifically, defining
\begin{align}
^{1}D = \sum_{ij}\;^{1}D_{j}^{i}\vert i \rangle \langle j \vert \\
^{2}D = \sum_{pq,rs}\;^{2}D_{rs}^{pq}\vert p q \rangle \langle r s \vert 
\end{align}
where
\begin{align}
^{1}D_{j}^{i} = \mathrm{Tr}[a_{i}^{\dagger}a_{j}  \;^{N}D ] = \langle \psi \vert a_{i}^{\dagger}a_{j} \vert \psi \rangle \\
^{2}D_{rs}^{pq} = \mathrm{Tr}[ a_{p}^{\dagger}a_{q}^{\dagger}a_{s}a_{r} \;^{N}D ] = \langle \psi \vert  a_{p}^{\dagger}a_{q}^{\dagger}a_{s}a_{r} \vert \psi \rangle
\end{align}
the simplest of these constraints may be enumerated as:
\begin{enumerate}
\begin{item}
Hermiticity of the density matrices
\begin{align}
^{1}D_{i}^{j} = \left(\;^{1}D_{j}^{i}\right)^{*}\\
^{2}D_{rs}^{pq} = \left(\;^{2}D_{pq}^{rs}\right)^{*}\\
\end{align}
\end{item}
\begin{item}
Antisymmetry of the $2$-particle marginal
\begin{align}
^{2}D_{rs}^{pq} = - \;^{2}D_{sr}^{pq} = -\; ^{2}D_{rs}^{qp} = \;^{2}D_{sr}^{qp}
\end{align}
\end{item}
\begin{item}
The $(p-1)$-marginal is related to the $p$-marginal by contraction--e.g. the $2$-marginal can be contracted to the $1$-marginal
\begin{align}
^{1}D_{j}^{i} = \frac{1}{n-1}\sum_{k}\;^{2}D_{jk}^{ik}
\end{align}
\end{item}
\begin{item}
The trace of each marginal is fixed by the number of particles in the system
\begin{align}
\mathrm{Tr}[\;^{1}D] & = n\\
\mathrm{Tr}[\;^{2}D] & = n ( n -1 )
\end{align}
\end{item}
\begin{item}
The marginals are proportional to density matrices and are thus positive semidefinite
\begin{align}
\{{}^{1}D, {}^{2}D\} \succeq 0
\end{align}
\end{item}
\end{enumerate}
Additional constraints based on the quantum numbers of $S^{2}$ and $S^{z}$ operators can be derived for each marginal~\cite{alcoba2005spin}. A short description of the form of the linear constraints inspired by fixed particle number $\langle n \rangle$, fixed total angular momentum $\langle S^{2} \rangle$, and fixed projected total angular momentum $\langle S_{z} \rangle$ are described in Appendix~\ref{sec:rdm_observables}.

\section{The $n$-Representability problem}\label{sec:nrep}
In the previous section we provided a brief introduction to the notation and setup of problems formulated through their marginal distributions.  Here, we review a concise and elegant theoretical framework that allows one to derive the full set of representability conditions for fermionic systems, from which one may select a subset to form efficient approximations.  The polynomial size of the $p$-marginals makes them attractive candidates for use as the representation of quantum systems. This was originally noticed by Coulson and Coleman~\cite{lowdin1987some, RevModPhys.35.668} in the context of quantum chemical Hamiltonians where the energy operator is $2$-local and thus a linear functional of the $2$-RDM.  The characterization and structure of valid $2$-marginals arising from the integration of a fermionic density matrix led to the field of $n$-representability~\cite{RevModPhys.35.668, coleman2000reduced}.

A significant amount of progress has been made in tackling the $n$-representability problem; most notably, the original works by Erdahl~\cite{erdahl1978representability}, Percus and Garrod~\cite{percus1978role, garrod1964reduction}, and Mazziotti~\cite{mazziotti2002variational, mazziotti2001uncertainty, PhysRevA.94.032516, PhysRevLett.108.263002} formalize the $n$-representability problem by specification of an approximate set of constraints by parameterizing the polar cone of the set of $2$-RDMs. Recently, Mazziotti formalized the complete structure of the ensemble $n$-representability constraints~\cite{PhysRevLett.108.263002} and gave the structure of pure-state constraints for states with fixed particle number~\cite{PhysRevA.94.032516}. $n$-Representability has also been greatly influenced by quantum information:  Liu~\cite{liu2006consistency} demonstrated that the $n$-representability problem is QMA-complete and Bravyi~\cite{bravyi2003requirements} and Klyacho~\cite{klyachko2006quantum} enumerated $n$-representability constraints for marginals of a pure-state.  

\subsection{$n$-Representability by Characterizing the Polar Cone}
We begin with a geometric picture of the constraints within the space of fermionic density matrices.  The formal characterization of the $n$-representable set of $^{2}D$ operators relies on characterizing the polar cone of the $2$-marginals~\cite{coleman1980reduced, coleman1978reduced, coleman2000reduced}.  Consider the convex set of $2$-marginals ${}^{2}\mathcal{D}$ acting on the anti-symmetric two-fermion space $\wedge^{2}\mathcal{H}$, which may be defined in terms of its basis vectors $\{a \wedge b = a \otimes b - b \otimes a\}$ for $a, b \in \mathcal{H}$.  The polar cone is defined as the subset of Hermitian operators that satisfy the positive projection condition,
\begin{align}\label{eq:polar_cone}
{}^{2}\tilde{\mathcal{D}} = \{ {}^{2}B \in \wedge^{2}\mathcal{H} \vert \langle {}^{2}B \vert {}^{2}D\rangle \geq 0  \; \forall \;{}^{2}D \in \;{}^{2}\mathcal{D} \}.
\end{align}
Operators of the polar cone ${}^{2}B \in {}^{2}\tilde{\mathcal{D}}$ are positive operators with respect to the $2$-RDM, which implies their non-negativity with respect to all fermionic density matrices $^{N}D$ when lifted to the $n$-particle space. In quantum information this lifting procedures is accomplished by taking the tensor product of the operator with identities. For fermions, the lifting procedure involves the tensor product with the appropriate antisymmeterization operations.

The bipolar theorem states that elements of ${}^{2}\mathcal{D}$ are completely characterized by the polar cone ${}^{2}\tilde{\mathcal{D}}$
\begin{align}
{}^{2}\mathcal{D} = \{ {}^{2}D \in \wedge^{2}\mathcal{H} \vert \langle {}^{2}B \vert {}^{2}D \rangle \geq 0 \; \forall \;  {}^{2}B \in {}^{2}\tilde{\mathcal{D}} \}.
\end{align}
Though specification of inequalities with elements of the polar cone provides a characterization of ${}^{2}\mathcal{D}$, it has the major drawbacks that a) there are infinitely many possible ${}^{2}B$ operators to check and b) checking ${}^{2}B$ requires checking if an exponentially large operator is positive~\cite{coleman1977convex}.  

The key approximation within $n$-representability is constructing a polynomial size approximation to the polar cone ${}^{2}\tilde{\mathcal{D}}_{a}$ and deriving conditions on the $2$-marginal through Eq~\eqref{eq:polar_cone}.  This is achieved by selecting a $k$th-order (where $k < n$) operator basis for ${}^{2}\tilde{\mathcal{D}}_{a}$.  Given that the ${}^{2}\tilde{\mathcal{D}}_{a} \subset {}^{2}\tilde{\mathcal{D}}$ any representability conditions derived from ${}^{2}\tilde{\mathcal{D}}_{a}$ will be an approximate set of representability conditions. By duality, the polar of ${}^{2}\tilde{\mathcal{D}}_{a}$ implies ${}^{2}\mathcal{D} \subset {}^{2}\mathcal{D}_{a}$.  This naturally explains why variational calculations using the reduced-density matrix and approximate $n$-representability constraints are strictly a lower bound to the true energy~\cite{baumgratz2012lower, fosso2016accuracy, nakata2001variational, zhao2004reduced, mazziotti2001uncertainty, rubin2015strong, hammond2005variational, mazziotti2011large}.

In the $n$-representability literature related to simulating fermions, fermionic operators up to a particular order--e.g. $\{a_{i}, a_{i}^{\dagger}, a_{i}a_{j}, a_{i}^{\dagger}a_{j}^{\dagger}, \dots \}$--are used as the operator basis for the approximate polar cone.  Given a rank-$k$ operator basis for the polar cone, we can define a real linear space of Hermitian operators $O_{k}^{\dagger}O_{k}$ where
\begin{align}\label{op_basis}
O_{k} = \sum_{k = 1}^{N} \prod_{j = 1, o\in \{1, \dagger\}}^{k} c_{k_{j}}\hat{a}_{k_{j}}^{o}
\end{align}
that when constrained to be non-negative (implied by Eq.~\eqref{eq:polar_cone}) form a necessary set of conditions on the $p$-marginals of the von Neumann density matrix.  

As an example, we will derive the famous $2$-positivity conditions by restricting the rank of the monomials in the operator basis to rank less than $2$ as
\begin{align}\label{eq:arbitrary_2op}
O_{2} = \sum_{i}c_{i}^{a}\hat{a}_{i} + \sum_{i}c_{i}^{b}\hat{a}_{i}^{\dagger} + \sum_{ij}c_{ij}^{c}\hat{a}_{i}\hat{a}_{j} + \sum_{ij}c_{ij}^{d}\hat{a}_{i}\hat{a}_{j}^{\dagger} + \sum_{ij}c_{ij}^{e}\hat{a}_{i}^{\dagger}\hat{a}_{j} + \sum_{ij}c_{ij}^{f}\hat{a}_{i}^{\dagger}\hat{a}_{j}^{\dagger},
\end{align}
setting
\begin{align}
M_{2} = O_{2}^{\dagger}O_{2},
\end{align}
and requiring that $M_{2} \succeq 0$.  The $\{c\}$ coefficients in Eq.~\eqref{eq:arbitrary_2op} specify an arbitrary element of the approximate polar cone $^{2}\tilde{\mathcal{D}}_{a}$ in a similar fashion to how a sum-of-squares polynomial can be expressed as a quadratic form $c^{T}Ac$ where elements of $A$ represent various products of monomials. Considering symmetries of the system, such as fixed particle number, reduces the large matrix $M_{2}$ to a block diagonal matrix~\cite{baumgratz2012lower}.  In this work we consider spinless fermionic Hamiltonians that commute with the number operator of our system and thus we can decompose $M_{2}$ into blocks where monomials correspond to number preserving operators--i.e $\{a_{i}^{\dagger}a_{j}, a_{j}^{\dagger}a_{i}, ...\}$.  Restricting the operator $M_{2}$ to be non-negative against the $1$-RDM and $2$-RDM for all values of $c$ yields the following inequalities
\begin{align}
\sum_{ij}c_{i}c_{j}^{*}\langle \psi \vert a_{j}^{\dagger}a_{i}\vert \psi \rangle  \geq 0\\
\sum_{ij}c_{i}c_{j}^{*}\langle \psi \vert a_{j}a_{i}^{\dagger} \vert \psi \rangle \geq 0 \\
\sum_{ij, kl}c_{ij}c_{kl}^{*}\langle \psi \vert a_{i}^{\dagger}a_{j}^{\dagger}a_{l}a_{k}\vert \psi \rangle \geq 0\\
\sum_{ij, kl}c_{ij}c_{kl}^{*}\langle \psi \vert a_{i}a_{j}a_{l}^{\dagger}a_{k}^{\dagger}\vert \psi \rangle \geq 0\\
\sum_{ij, kl}c_{ij}c_{kl}^{*}\langle \psi \vert a_{i}^{\dagger}a_{j}a_{l}^{\dagger}a_{k}\vert \psi \rangle \geq 0
\end{align}
where $\psi$ is an arbitrary state. These conditions imply that the following matrices are positive semidefinite
\begin{align}
{}^{1}D= \langle \psi \vert a_{j}^{\dagger}a_{i}\vert \psi \rangle  \succeq 0 \label{eq:opdm}\\
{}^{1}Q = \langle \psi \vert a_{j}a_{i}^{\dagger} \vert \psi \rangle \succeq 0 \label{eq:oqdm}\\
{}^{2}D = \langle \psi \vert a_{i}^{\dagger}a_{j}^{\dagger}a_{l}a_{k}\vert \psi \rangle \succeq 0 \label{eq:tpdm}\\
{}^{2}Q = \langle \psi \vert a_{i}a_{j}a_{l}^{\dagger}a_{k}^{\dagger}\vert \psi \rangle \succeq 0 \label{eq:tqdm}\\
{}^{2}G = \langle \psi \vert a_{i}^{\dagger}a_{j}a_{l}^{\dagger}a_{k}\vert \psi \rangle \succeq 0. \label{eq:phdm}
\end{align}
The positivity of these matrices canonically known as $\{ {}^{1}D, {}^{1}Q, {}^{2}D, {}^{2}Q, {}^{2}G \}$ form a set of necessary conditions that the $2$-marginal must obey.  Clearly, as the $2$-marginal is included in our set, the positivity of this operator naturally appears when building constraints starting from a polar cone picture.  Though these constraints are formulated with pure states it is simple to show these conditions hold for mixed states as well.

The positivity of the operators on $\mathcal{H}$--$\{{}^{1}D, {}^{1}Q\}$--and $\wedge^{2}\mathcal{H}$--$\{{}^{2}D, {}^{2}Q, {}^{2}G\}$--are constrained to live in the space defined by equalities obtained by rearranging the fermionic ladder operators according to their anticommutation rules.  A full enumeration of the equality constraints is contained in Appendix~\ref{appendix:mapping_conditions}.  In this work we use the positivity constraints and the linear constraints from the anticommutation relationships as a set of constraints that $2$-RDMs measured from a quantum resource must satisfy.  This enables us to enhance the accuracy of estimation of quantities through employing basic physical relations.  

\section{Concentration of measure in $p$-RDMS}\label{sec:concentration}
Hybrid quantum-classical schemes depend both on the ability to perform the partial tomography that has been discussed as well as some parameterization of the quantum state space.  Recent experimental proposals have considered the use of quantum states that are constructed from unitaries that are uniformly random with respect to the Haar measure acted upon a well defined initial state in order to demonstrate so-called ``quantum supremacy'' over classical devices~\cite{Boixo:2016}.  A natural question is to ask whether these states can be harnessed as a resource within hybrid schemes.  However, while these states are highly entangled, they demonstrate a number of surprising properties related to results on concentration of measure in high dimensional spaces.  From the discussion above, we know it is sufficient to characterize a local fermionic system by its reduced marginals, so we investigate these states in that setting.  In a colloquial sense, concentration results for random quantum states show that for many local observables, typical measurement results will yield the average value with overwhelming probability.  Here, specifically we investigate the implications of this for $p$-particle reduced density matrices.

A key result we will leverage from the theory of measure concentration is Levy's lemma~\cite{milman2009asymptotic}, which relates to the concentration of Lipshitz-continuous functions.  Levy's lemma is as follows: consider a Lipschitz-continuous function $f: S^{(2n -1)} \rightarrow \mathcal{R}$ with Lipschitz constant $C$, i.e. $|f(x) - f(y)| < C ||x - y||$ for all $x, y \in S^{(2n - 1)}$ where $||.||$ is the Euclidean norm in the surrounding space $\mathcal{R}^{2n} \supset S^{(2n-1)}$, and $S^{(2n-1)}$ is the unit sphere in $\mathcal{R}^{2n}$ which would correspond to a quantum state of $\text{log}_2 (n)$ qubits.  Drawing a point $x \in S^{(2n - 1)}$ at random with respect to the uniform measure on the sphere yields
\begin{align}
    \text{Prob}[|f(x) - \langle f \rangle| \geq \epsilon] \leq 2 \ \exp \left(- \frac{n \epsilon^2}{9 \pi^3 C^2} \right).
\end{align}
We will leverage this lemma together with the fact that the expectation value on a normalized quantum state of any Hermitian operator $A$ with bounded spectrum is Lipschitz continuous, with a Lipschitz constant that may be bounded by the norm of the operator. 

Consider first the case of $1$-RDMs on a space consisting of any number of particles between $0$ and $M$ in $M$ spin-orbitals.  We are interested in the average value of a matrix element $[{}^1 D]_{ij} = \bra{\psi} a_i^\dagger a_j \ket{\psi}$ where $\ket{\psi}$ is a randomly selected pure state.  These states may be represented by the density matrix of all possible equally likely occupations,
\begin{align}
\rho = \frac{I}{d}
\end{align}
where $I$ is the identity matrix on the space of $2^M$ possible occupations and $d$ is the dimension of this matrix, $d=2^n$.  To evaluate a trace in this case, it suffices to choose the basis of all determinants ranging from $0$ to $M$ occupied spin-orbitals, created in a standardized ordering $\ket{\psi}_{S} = \prod_{k \in S} a_k^\dagger \ket{\ }$ where $\ket{\ }$ is the standard fermi vacuum state. To compute the average value of this operator, we thus need to evaluate
\begin{align}
\langle a_{i}^{\dagger} a_{j} \rangle = \mathrm{Tr}[\rho a_{i}^{\dagger} a_{j} ].
\end{align}
On expanding this trace in the determinant basis, one realizes that the terms vanish unless $i=j$, unless this index appears in the wavefunction, and there is at least $1$ particle in the wavefunction.  The trace is given by
\begin{align}
\langle a_i^\dagger a_j \rangle &= \frac{\delta_{ij}}{2^M} \sum_{n=0}^M   \left( \begin{array}{c} M-1 \\ n-1 \end{array} \right) \notag \\
&= \frac{\delta_{ij}}{2^M} \sum_{n=0}^M \frac{n}{M}   \left( \begin{array}{c} M \\ n \end{array} \right) \notag \\
&= \frac{\delta_{ij}}{M 2^M} \left(M 2^{M-1} \right) \notag \\
&= \frac{\delta_{ij}}{2}.
\end{align}
Thus the average $1$-RDM is a diagonal matrix with entries $1/2$, corresponding to a system with an average number of particle of $M/2$.  Note that for the operator $a_i^\dagger a_j$, the Lipschitz constant can be safely bounded by $1$, as independent of the distance in space, this expectation value can differ by at most $1$ due to the spectrum of the operator.  As a result, Levy's lemma informs us how large we expect typical deviations to be from this average matrix element as
\begin{align}
    \text{Prob}[\ |\bra{\psi}a_i^\dagger a_j\ket{\psi} - \langle a_i^\dagger a_j \rangle | \geq \epsilon] \leq 2 \ \exp \left(- \frac{2^M \epsilon^2}{9 \pi^3} \right).
\end{align}
Examining the case of $2$-RDMs now using the same techniques, we find that
\begin{align}
    [{}^2 D]^{ij}_{kl} &=  \frac{1}{2} \langle a_i^\dagger a_j^\dagger a_l a_k \rangle = \frac{1}{2^{M+1}} \left(\delta_{ik} \delta_{jl} - \delta_{il}\delta_{jk} \right)
    \left(1 - \delta_{ij} \right) \left( 1 - \delta_{kl} \right) \sum_{n=0}^M \left( \begin{array}{c} M-2 \\ n-2 \end{array} \right) \notag \\
    &= \frac{1}{2^{M+1} M (M-1)} \left(\delta_{ik} \delta_{jl} - \delta_{il}\delta_{jk} \right)
    \left(1 - \delta_{ij} \right) \left( 1 - \delta_{kl} \right) \sum_{n=0}^M (n^2 - n) \left( \begin{array}{c} M \\ n \end{array} \right) \notag \\
    &= \frac{1}{2^{M+1} M (M-1)} \left(\delta_{ik} \delta_{jl} - \delta_{il}\delta_{jk} \right)
    \left(1 - \delta_{ij} \right) \left( 1 - \delta_{kl} \right) \left( (M+M^2)2^{M - 2} - M 2^{M-1} \right) \notag \\
    &= \frac{1}{8} \left(\delta_{ik} \delta_{jl} - \delta_{il}\delta_{jk} \right)
    \left(1 - \delta_{ij} \right) \left( 1 - \delta_{kl} \right).
\end{align}
From this, we see that the $2-$RDM average also represents a generalization of a diagonal matrix to a 4-tensor with signs reflecting the anti-symmetry properties of the electrons.  Moreover, it is easy to see from the operator $\frac{1}{2} a_i^\dagger a_j^\dagger a_l a_k$ that the same concentration results hold for each individual matrix element as in the $1$-RDM case, except that the Lipzschitz constant is modified by the normalization factor $1/2$.  Using the same process, one may derive similar results for higher particle reduced density matrices, and conclude that all higher RDMs concentrate towards diagonal matrices at a similar rate, with modified Lipschitz constants due to normalization.  

Now we consider the case where one restricts to random states within an $n$-particle subspace.  Such a state can be represented by a density matrix of the form
\begin{align}
    \rho_R = \frac{I_R}{d_R}
\end{align}
where $I_R$ is the identity operator on the full space of $2^M$ spin-orbitals subject to the restriction $R$ to the space of $n$ particles, and $d$ is the dimension of that space.  
To evaluate this trace, we consider the basis of $n$-particle determinants.  Using the same machinery as above, but restricting the sum to the case of only $n$-particles, we find
\begin{align}
\langle a_i^\dagger a_j \rangle_{R} &= \frac{\delta_{ij}}{d_R} \left( \begin{array}{c} M-1 \\ n-1 \end{array} \right) \notag \\
&= \delta_{ij} \left( \begin{array}{c} M-1 \\ n-1 \end{array} \right) \left( \begin{array}{c} M \\ n \end{array} \right)^{-1} \notag \\
&= \frac{n}{M}.
\end{align}
Thus the average $1$-RDM in a space of randomly generated $n$-particle states in $M$ orbitals is the diagonal matrix with equally probable occupations on all sites. The exact convergence results of Levy's lemma depend on the spherical geometry of states, so direct application of the concentration results would require a modification of the lemma.  However, we may embed the allowable quantum states into the space of $\lceil \text{log}_2(d_R) \rceil$ qubits and leveraging the fact that the norm of the operator of interest, and thus Lipschitz constant will remain the same.  If one generates random quantum states within this embedded space, we can see that the concentration result holds with the same Lipschitz constant but modified dimension $\left( \begin{array}{c} M \\ n \end{array} \right)$.

A consequence of these results is that random quantum states generate $p$-particle marginals that are effectively trivial, concentrating exponentially quickly to their average value as a function of system size.  This would mean that one can evaluate the expectation value to specified precision of any $p$-particle observable (where $p$ is held fixed as system size grows) efficiently on a classical computer for a random quantum state. Thus, we conclude meaningful explorations of the space of quantum states must be structured, whether it be in the path of time evolution or the design of a variational ansatz.  If an ansatz and method used for hybrid quantum-classical methods cannot easily exit the space of Haar random states, the above analysis dictates it is doomed to give trivial observables for fermionic systems at relatively small system sizes.

\section{Reducing Operator Sample Variance Using $n$-Representability Constraints}\label{sec:op_av_bound}

\subsection{Optimal Operator Averaging}\label{sec:opt_op_average}
Any $L$-sparse Hermitian operator on Hilbert space can be expressed as
\begin{equation}
\label{eq:local_op}
H = \sum_{\ell=0}^{L-1} w_\ell H_\ell
\quad \quad \quad
\textrm{s.t.} \quad w_\ell \in \mathbb{R}
\quad \quad \quad
H_\ell^2 = \openone
\end{equation}
where $w_\ell$ are real scalars and $H_\ell$ are 1-sparse self-inverse operators which act on qubits.  In second quantized formulations of electronic structure, the $H_\ell$ are typically a special case of 1-sparse operators that have particularly convenient properties for measurement, namely they are strings of Pauli operators. Very often one is interested in estimating $\avg{H}$ by making measurements on $M$ independent copies of a state $\ket{\psi}$. For instance, the typical procedure in variational algorithms is to estimate the energy $\avg{H}$ by repeatedly preparing a state and performing projective measurements onto the eigenstates of Pauli operators $\avg{H_\ell}$. Since the $H_\ell$ are self-inverse, the intrinsic variance of these projective measurements is computed as
\begin{equation}
\label{eq:sigmas}
\sigma_\ell^2 = \avg{H_\ell^2} - \avg{H_\ell}^2 = 1 - \avg{H_\ell}^2 \leq 1.
\end{equation}
As sample variance is due to statistical fluctuations which are uncorrelated from sample to sample, the total variance of $\avg{H}$ scales as
\begin{equation}
\label{eq:error_bound}
\epsilon = \sqrt{\sum_{\ell=0}^{L-1} \frac{w_\ell^2 \sigma_\ell^2}{M_\ell}} = \sqrt{\sum_{\ell=0}^{L-1} w_\ell^2 \left(\frac{1 - \avg{H_\ell}^2}{M_\ell}\right)}.
\end{equation}
The real question is how to choose the number of samples for each term in the Hamiltonian $M_\ell$ in order to minimize $\epsilon$ for the fewest overall measurements $M = \sum_\ell M_\ell$.

In \cite{Wecker2015a}, it is suggested that one choose $M_\ell \propto |w_\ell|$ with no guarantee of optimality. Here we prove that this choice is optimal by application of the Lagrange conditions  (no proof was provided in \cite{Wecker2015a}). We start with the Lagrangian
\begin{equation}
{\cal L} = \sum_{\ell=0}^{L-1} M_\ell + \lambda \left(\sum_{\ell=0}^{L-1} \frac{w_\ell^2 \sigma_\ell^2}{M_\ell} - \epsilon^2\right)
\end{equation}
where the constant $\lambda$ is the Lagrangian multiplier. Our goal will be to solve the following expression for $M_\ell$,
\begin{equation}
\min_{M_\ell} \max_\lambda {\cal L} = \min_{M_\ell} M.
\end{equation}
Accordingly, we take the derivative of ${\cal L}$ with respect to $M_\ell$ to find,
\begin{equation}
\frac{\partial {\cal L}}{\partial M_\ell} = \sum_{\ell=0}^{L-1} \left(1 - \lambda \frac{w_\ell^2 \sigma_\ell^2}{M_\ell^2}\right) = 0 \quad \quad \rightarrow \quad \quad M_\ell = \sqrt{\lambda} | w_\ell | \sigma_\ell.
\end{equation}
Plugging this back into \eq{error_bound}, we find exactly that
\begin{equation}
\epsilon^2 = \frac{1}{\sqrt{\lambda}} \sum_{\ell = 0}^{L-1} | w_\ell | \sigma_\ell \quad \quad \rightarrow \quad \quad \sqrt{\lambda} = \frac{1}{\epsilon^2} \sum_{\ell = 0}^{L-1} | w_\ell | \sigma_\ell.
\end{equation}
Therefore,
\begin{equation}
\label{eq:lambda_bound}
M = \sum_{\ell = 0}^{L-1} M_\ell = \left(\frac{1}{\epsilon} \sum_{\ell = 0}^{L-1} |w_\ell| \sigma_\ell \right)^2 \leq \frac{\Lambda^2}{\epsilon^2}
\quad \quad \quad
\Lambda = \sum_{\ell=0}^{L-1} \left | w_\ell \right |.
\end{equation}
If we insist on getting an asymptotic bound then we assume $\sigma_\ell = {\cal O}(1)$ and this leads us to confirm the optimality of the suggestion of \cite{Wecker2015a}. 


\subsection{Reducing Variance Using $n$-Representability}\label{sec:var_reduction}

\begin{figure}[b]
  \begin{subfigure}[b]{0.485\textwidth}
    \includegraphics[width=\textwidth]{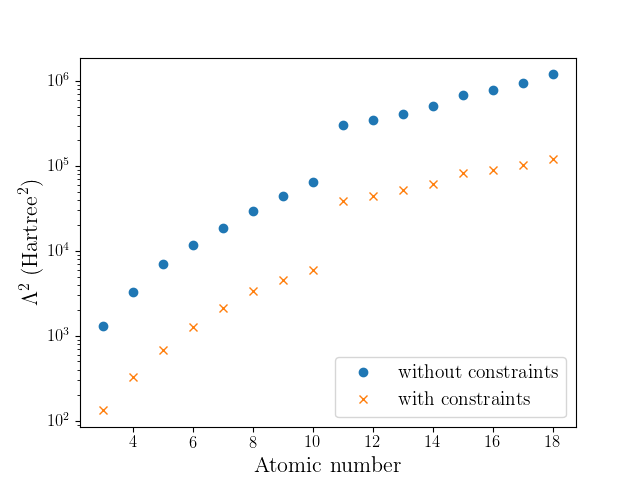}
    \caption{}
    \label{fig:periodic}
  \end{subfigure}
  \begin{subfigure}[b]{0.485\textwidth}
    \includegraphics[width=\textwidth]{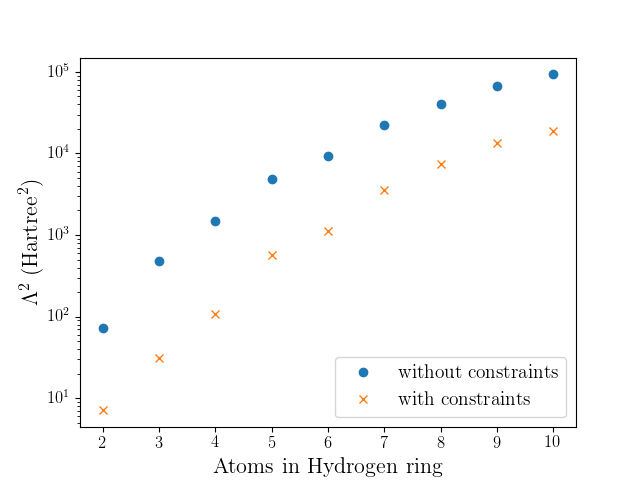}
    \caption{}
    \label{fig:rings}
  \end{subfigure}
  \begin{subfigure}[b]{0.485\textwidth}
    \includegraphics[width=\textwidth]{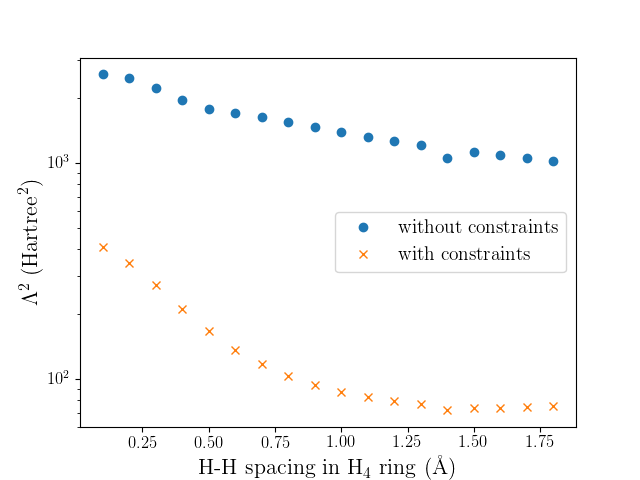}
    \caption{}
    \label{fig:geometry_trends}
  \end{subfigure}
    \begin{subfigure}[b]{0.485\textwidth}
    \includegraphics[width=\textwidth]{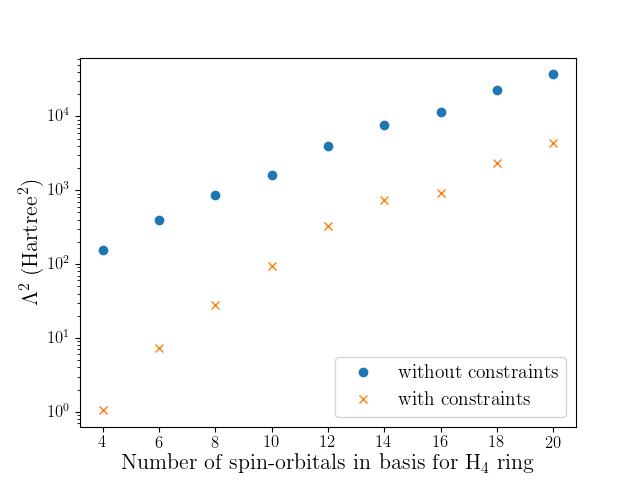}
    \caption{}
    \label{fig:basis_trends}
  \end{subfigure}
  \caption{\label{fig:lagrange_data} A series of plots showing values of $\Lambda^2$ and $\widetilde{\Lambda}^2$ as defined in \eq{lambda_bound} and \eq{reduced_measures}. In all plots, the blue circles correspond to the value of $\Lambda^2$ prior to applying the techniques of this section and the orange crosses correspond to the value of $\widetilde{\Lambda}^2$ after applying the techniques of this section. In \fig{periodic} we demonstrate our technique on single atom calculations in the minimal basis. We see a consistent improvement of about one order of magnitude with a jump in values between the second and third rows of the periodic table. In \fig{rings} we show a progression of hydrogen rings in the minimal basis of increased size where the distance between adjacent hydrogen atoms is fixed at the $\textrm{H}_2$ bond length of 0.7414 \AA. In \fig{geometry_trends} we show how geometry affects these techniques by studying a square $\textrm{H}_4$ ring in the minimal basis as the spacing between hydrogens in the square is changed from 0.1 \AA \ to 1.8 \AA. Finally, in \fig{basis_trends} we examine how these techniques are effected as one increases the active space of an $\textrm{H}_4$ ring with atom spacing of 0.7414 \AA \ from four spin-orbitals to twenty spin-orbitals with calculations performed in a double zeta (cc-pVDZ) basis.}
\end{figure}

After mapping RDM elements to Pauli matrices (e.g. by using fermionic transforms such as the Jordan-Wigner or Bravyi-Kitaev transformation), one can express all equality $n$-representability constraints in the notation of \eq{local_op}. Assuming a list of  $K$  equality constraints, we will express the $k^\textrm{th}$ constraint $C_k$ as 
\begin{equation}
\label{eq:constraints}
C_k = \sum_{\ell=0}^{L-1} c_{k,\ell} \avg{H_\ell} = 0
\quad \quad \quad
c_{k,\ell} \in \mathbb{R}
\end{equation}
where we can always choose to have a constant term in the set (e.g. $H_0 = \openone$) so that the equality sums to zero\footnote{Throughout this section, the $H_\ell$ should be the full set of terms that map to qubits from the 2-RDM. Terms that do not appear in the operator of interest (for instance, 2-RDM elements that are not in the Hamiltonian) simply have a coefficient of zero.}. These constraints provide extra information about the relationships between expectation values which, in principle, should allow us to make fewer measurements. One very straightforward way to exploit this information in order to make fewer measurements is to add these constraints to the operator of interest in order to minimize the associated $\Lambda$ from \eq{lambda_bound}. Specifically, we have that
\begin{equation}
\label{eq:constraint_ham}
\widetilde{H} = H + \sum_{k=0}^{K-1} \beta_k C_k = \sum_{\ell=0}^{L-1} \left(w_\ell + \sum_{k=0}^{K-1} \beta_k c_{k,\ell}\right) H_\ell
\quad \quad \quad
\avg{H} = \avg{\widetilde{H}}
\quad \forall \quad \beta_k \in \mathbb{R}
\end{equation}
where the relation $\avg{H} = \avg{\widetilde{H}}$ follows from the observation that $C_k = 0$ for $n$-representable states due to the definition of $C_k$ in \eq{constraints}. When we do this, from \eq{constraint_ham} and \eq{lambda_bound} we can see that the number of measurements required is expected to scale as
\begin{equation}
\label{eq:reduced_measures}
\widetilde{M} = \left(\frac{1}{\epsilon} \sum_{\ell=0}^{L-1} \left(w_\ell + \sum_{k=0}^{K-1} \beta_k c_{k,\ell}\right) \right)^2 \leq \frac{\widetilde{\Lambda}^2}{\epsilon^2}
\quad \quad \quad
\widetilde{\Lambda} = \sum_{\ell=0}^{L-1} \left |w_\ell + \sum_{k=0}^{K-1} \beta_k c_{k,\ell}\right|.
\end{equation}
In order to minimize measurements then, the strategy is to compute
\begin{equation}
\beta^* = \argmin_{\beta} \left(\sum_{\ell=0}^{L-1} \left |w_\ell + \sum_{k=0}^{K-1} \beta_k c_{k,\ell} \right |\right)
\quad \quad \textrm{or} \quad \quad
\beta^* = \argmin_{\beta} \left(\widetilde{\Lambda}\right),
\end{equation}
depending on whether or not one has any meaningful prior on the expectation values $\avg{H_\ell}$ (which would provide a meaningful prior on $\sigma_\ell$ via \eq{sigmas}).

We can easily recast this optimization problem in a form amenable to efficient solution by common numerical methods. To do this, we think of the original Hamiltonian $H$ as being expressed as a vector $v_H$ where each element of a vector represents a different fermionic operator; for example, we could map term $a^\dagger_p a_q$ to vector element $1 + p + q \, N$ and map $a^\dagger_p a^\dagger_q a_r a_s$ to $1 + N^2 + p + q \, N + r \, N^2 + s \, N^3$. The coefficients of the vector correspond to the coefficients of the term. Likewise, we can represent all of the constraints in a matrix $C$ of dimension $K \times L$ where each constraint $C_k$ is a row of the matrix vectorized in the same way as $v_H$. Then, we see that the optimization task at hand can be expressed as
\begin{equation}
\beta^* = \argmin_\beta \left \| v_H - C^\top \beta \right \|_1
\end{equation}
where $\beta$ is a vector of dimension $K$. We can see now that this is a convex $L_1$ minimization. Such minimizations can be solved efficiently using simplex methods. We can cast $L_1$ minimization as the linear program:
\begin{equation}
\textrm{minimize} \quad \openone^{\top} q
\quad \quad \quad
\textrm{subject to} \quad -q \leq v_H - C^\top \beta \leq q,
\end{equation}
where $q$ is an auxiliary vector variable. We provide freely available source code that generates the equality constraints and performs this optimization in the open source project OpenFermion \cite{OpenFermion}. Our code uses GLPK (GNU Linear Programming Kit) for the linear programming component via a Python wrapper known as CVXOPT. We show results of several numerical experiments which demonstrate the effectiveness of our method in \fig{lagrange_data}. These numerical experiments involved linear $n$-representability constraints coming from the $1$-RDM trace, $1$-RDM Hermiticity, $2$-RDM trace, $2$-RDM Hermiticity, $2$-RDM to $1$-RDM contraction, and the mappings between the $2$-RDM and the other marginals in the $2$-positive set.  The set of mappings are described in Appendix~\ref{appendix:mapping_conditions}.  These techniques often reduce the required measurements by an order of magnitude or more.

Note that the method discussed here is actually quite a bit more general than presented. In particular, we have found an interesting method for transforming the Hamiltonian in a way that leaves its spectrum invariant in the fixed particle number sector. One might postulate that this method could also be used to optimize other simulation metrics, for instance, to reduce Trotter errors which are well known to be related to the norm of the Hamiltonian. Note that all constraints $C_k$ will be either Hermitian or anti-Hermitian operators. In particular, the Hermiticity constraints take the form of constraining anti-Hermitian components of the density matrix to be zero (thus those $C_k$ are themselves anti-Hermitian operators). So after applying the procedure here, one may end up with a $\widetilde{H}$ that is not Hermitian. Fortunately, one can restore Hermiticity without changing the value of $\widetilde{\Lambda}$ by creating a new Hamiltonian, $H^\star = (\widetilde{H} + \widetilde{H}^\dagger) / 2$. $H^\star$ will be isospectral to $H$ in the $n$-electron manifold and will have the same $\widetilde{\Lambda}$ as $\widetilde{H}$.

It remains an open question if this variance reduction technique can be applied to other hybrid algorithms such as QAOA or quantum spin Hamiltonians.  In the QAOA case, spin-marginals with significantly less structure than fermionic marginals must be considered.  Linear constraints outside of the consistency of the marginals with overlapping support are generally unknown for an arbitrary Hamiltonian encoding a combinatorial optimization problem as is done in QAOA.  For spin Hamiltonians constraints generated by fixed values of $\langle S^{2} \rangle$ can also be considered. Further investigation of the consistency constraints and eigenvalue constraints in the form of pure-state constraints may provide additional variance reduction for problems described by fermionic and spin Hamiltonians.

\section{$n$-Representability informed projection of $2$-RDMs}\label{sec:nrep_proj}
In this section we discuss the possibility of using $n$-representability conditions to improve $2$-RDMs sampled from a quantum device.  Errors in the $2$-RDM measured from a quantum state can appear in multiple ways: 1) stochastic errors associated with the operator averaging techniques used to measure expected values and 2) device errors such as unexpected measurement correlations.  We explored the utility of $2$-marginal reconstruction schemes using $n$-representability rules to remove stochastic errors associated with sampling and state errors corresponding to noise of the device corrupting the intended state.

The general strategy is to design a process that projects a $2$-RDM back into the set of $n$-representable $2$-RDMs while balancing data-collection time and classical post-processing time.  In this section we first discuss two simple purification procedures: positive-semidefinite projection of the measured $2$-RDM with and without fixed-trace.  These simple projection techniques are compared against procedures involving projections with knowledge of representability constraints.
\subsection{Positive-Semidefinite Projection and Positive-Semidefinite Projection with Fixed Trace}
The simplest of the $n$-representability rules enforce the $2$-RDM to be Hermitian and non-negative with fixed trace.  Given a measured $2$-RDM we can define a computational procedure that determines the closest positive-semidefinite matrix.
\begin{align}\label{eq:marginal_opt}
\mathrm{min}\;\; ||^{2}D - \;^{2}D_{\mathrm{measured}}||_{2} \\
\mathrm{s.t.}\;\;^{2}D \succeq 0 \;\; \mathrm{Tr}[\;^{2}D] = n (n - 1)
\end{align}
The normalization is fixed by the particle number of the system. Without the trace condition the $2$-RDM that minimizes the objective in Eq.~\eqref{eq:marginal_opt} is the marginal constructed from the non-negative eigenvalues and eigenvectors of $^{2}D_{\mathrm{measured}}$~\cite{higham1988computing}.  The procedure for finding a fixed-trace positive-semidefinite projection have appeared in contexts such as tomography ~\cite{PhysRevLett.108.070502}, iterative purification of $2$-RDMs from response theory~\cite{lanssens2017method}, and finding positive-semidefinite correlation matrices~\cite{higham2002computing, higham1988computing}.  This projection procedure benefits from computational simplicity but suffers from the lack of information about representability conditions.  Therefore, given a sufficiently corrupted $^{2}D_{\mathrm{measured}}$, physicality is not guaranteed after projection.
\subsection{RDM Reconstruction with Approximate Representability Constraints}\label{sec:reconstruct}
In order improve the projection criteria we add additional $n$-representability constraints to the minimization procedure outlined in Eq.~\eqref{eq:marginal_opt}.   Given a collection of $2$-RDM elements at some unknown precision, or possibly missing crucial elements, our reconstruction scheme seeks to minimize the Frobenius norm of the difference between the reconstructed $2$-RDM and the set of known measurements subject to approximate $n$-representability constraints.  Denoting $E$ to be the difference between the reconstructed $2$-RDM and $^{2}D_{\mathrm{measured}}$ the minimization procedure can be formulated as the following non-convex optimization problem:
\begin{align}\label{nonconvex}
& \text{min} ||E||_{F}^{2}\\
& \text{s.t.} \text{Tr}[\;^{2}D] = n (n - 1) \nonumber \\
& \{ \;^{1}D, \;^{1}Q, \;^{2}D,\; ^{2}Q, \; ^{2}G \} \succeq 0 \nonumber \\
& A_{1}(^{1}D) \rightarrow \;^{1}Q \;\;,\;\; A_{2}(^{2}D) \rightarrow \;^{1}D \nonumber \\
& A_{3}(^{2}D) \rightarrow \;^{2}Q \;\;, \;\;A_{4}(^{2}D) \rightarrow \;^{1}G  \nonumber \\
\end{align}
where $A_{i}$ is the map from one marginal to another required by the fermionic ladder operator algebra.  The details of these mappings can be found in Appendix~\ref{appendix:rdms_struct}. The squared Frobenius norm of the error $||E||_{F}$ is quadratic in $2$-RDM. The optimization problem specified in Eq.~\eqref{nonconvex} can be relaxed to a semidefinite program (SDP) by taking the Schur complement in the identity block of the large matrix $M$
\begin{align}
M =
\begin{pmatrix}
I & E \\
E^{\dagger} & F
\end{pmatrix}
\succeq 0
\end{align}
constrained to be positive-semidefinite. In $M$, $I$ is the identity matrix, $F$ is a matrix of free variables, and $E$ is the error between the reconstructed $^{2}D$ and the $^{2}D_{\mathrm{measured}}$.  Taking the Schur complement in the identity block of $M$ gives
\begin{align}
F - E^{\dagger}E \succeq 0
\end{align}
or 
\begin{align}\label{rearrangedME}
F \succeq  E^{\dagger}E.
\end{align}
Noting that the Frobenius norm of a matrix $A$, $||A||_{F}$, is given as $\sqrt{\text{Tr}[A^{\dagger}A]}$, taking the trace of Eq.~\eqref{rearrangedME} gives the semidefinite relaxation of minimizing the Frobenius norm
\begin{align}
\text{Tr}[F] \succeq  \text{Tr}[E^{\dagger}E]
\end{align}
\begin{align}
||E||_{F}^{2} = \text{Tr}[E^{\dagger}E].
\end{align}
 We can now formulate the non-convex RDM reconstruction scheme in Eq.~\eqref{nonconvex} in terms of a semidefinite program:  
\begin{align}\label{eq:sdp_prog_reconstruct}
& \text{min} \; \text{Tr}[F] \\
& \text{s.t. } \{^{1}D, ^{1}Q, ^{2}D, ^{2}Q, ^{2}G, M \} \succeq 0 \nonumber \\
& A_{1}(^{1}D) \rightarrow \;^{1}Q \;\;,\;\; A_{2}(^{2}D) \rightarrow \;^{1}D \nonumber \\
& A_{3}(^{2}D) \rightarrow \;^{2}Q \;\;, \;\;A_{4}(^{2}D) \rightarrow \;^{1}G  \nonumber \\
& A_{5}(^{2}D) \rightarrow \; M\;\;, \;\;A_{6}(^{2}D) \rightarrow n(n - 1)\nonumber \\
& A_{7}(^{2}D) \rightarrow \; ^{2}D
\end{align}
where $A_{i}$ are the linear maps between $^{2}D$ and the other matrices $\{^{1}D, ^{1}Q, ^{2}D, ^{2}Q, ^{2}G, M \}$ along with the trace constraint and antisymmetry constraint on $^{2}D$.  These maps are described explicitly in Appendix~\ref{appendix:mapping_conditions} and are used in the Section~\ref{sec:results} for the SDP reconstruction program.
\subsection{Iterative Procedure for Projecting Noisy $2$-RDMs Into the Approximate $n$-Representable Subspace}
Although the SDP projection procedure can be extended to include better approximate $n$-representability conditions, it suffers from the requirement of solving a semidefinite program.  Despite the fact that an SDP can be solved in polynomial time with respect to the total number of variables and constraints, the high-order polynomial scaling of SDP algorithms makes the SDP-based project method infeasible for on-the-fly or online projections.  An alternative to the SDP projection combines the faster projection techniques based on fixed-trace positive projection with augmented $n$-representability conditions.  The projection technique is an iterative procedure that was originally developed to enforce approximate $n$-representability on $2$-RDMs obtained through a response formalism\cite{lanssens2017method}.  
The algorithm involves sequentially mapping $^{2}D$ to $^{2}Q$ to $^{2}G$ and enforcing the positivity and trace constraints at each of the operators.  The algorithm's main drawback is that any representable $2$-RDM is a valid fixed point.  As a result, linear constraints on the $2$-RDM preserving projected spin and total spin expectation values are not enforced and there is no guarantee that the ${}^{2}D$ obtained from the iterative procedure is any closer to the true $2$-RDM.  Therefore, it is likely required that the input $2$-RDM measured from the quantum resources is sufficiently close to the true $2$-RDM for this procedure to be most successful.

\begin{figure}[ht]
    \centering
    \includegraphics[width=16cm]{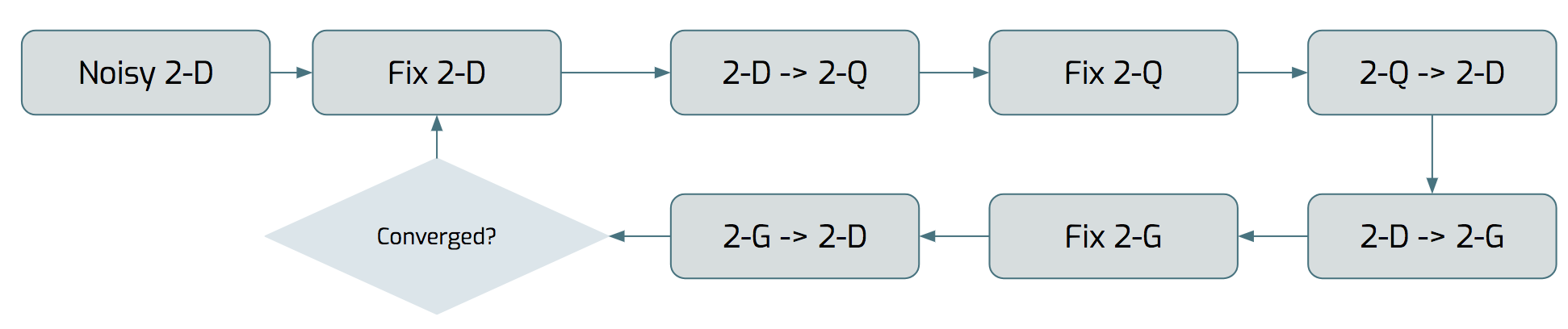}
    \caption{The iterative procedure for $2$-positive approximate $n$-representability constraints.  Starting with a noisy $2$-RDM the flow diagram is followed until the largest non-negative eigenvalue falls below a set threshold.  Eigenvalues are considered converged when the absolute value of the largest negative eigenvalue is less than  $1.0\times10^{-7}$. This stopping criteria was used for all numerical experiments with the $2$-positive iterative scheme.}
    \label{fig:iterRDM}
\end{figure}

The algorithm starts by enforcing Hermiticity of the given $2$-RDM matrix by averaging
\begin{align}
^{2}D^{s} = \frac{1}{2}\left( \;^{2}D_{\mathrm{meas}} + \;^{2}D_{\mathrm{meas}}^{\dagger} \right),
\end{align}
followed by a positive projection with fixed trace according to the procedure in Reference~\cite{higham2002computing}.  A detailed description of the algorithm for fixed-trace positive projection can also be found in the appendix of~\cite{lanssens2017method}.  Given a system with $r$ spin orbitals, $n$ particles, and $\eta = r - n$ holes, the iterative projection algorithm is as follows:
\begin{enumerate}
    \item[1] Enforce Hermiticity of $^{2}D$ and project to positive set with trace $n (n - 1)$
    \item[2] Map $^{2}D$ to the ${}^{2}Q$
    \item[3] Enforce Hermiticity of $^{2}Q$ and project to positive set with trace $\eta (\eta - 1)$ where $\eta$ is the number of holes
    \item[4] Map $^{2}Q$ to ${}^{2}G$
    \item[5] Enforce Hermiticity of $^{2}G$ and project to the positive set with trace $n (\eta + 1)$
    \item[6] Check the stopping condition associated with fixed trace for $^{2}D$, $^{2}Q$, and $^{2}G$ and positivity of their eigenvalues
\end{enumerate}
The iterative procedure is considered converged when the largest negative eigenvalue of any marginal in the $2$-positive set is below a set threshold. The total algorithm is depicted in Fig.~\ref{fig:iterRDM}

\subsection{Reconstruction results}\label{sec:results}
\subsubsection{Reconstruction of small systems}
To probe the utility of the $n$-representability inspired reconstruction schemes, we examined the accuracy of the energy and chemical properties obtained from $2$-RDMs with simulated sampling noise for diatomic hydrogen and a linear four-hydrogen chain.  All experiments involved corrupting the elements of a pure-state $2$-RDM with Gaussian noise proportional to the amount of samples used in operator averaging, followed by reconstructing the corrupted marginal with the four projection procedures outline above. The accuracy and precision of the reconstructed energies, particle-number, projected spin expectation $\langle S_{z} \rangle$, and total spin $\langle S^{2} \rangle$ are compared to provide the noise tolerance and precision of the various reconstruction schemes.

We obtained the Hamiltonian and ground-state wavefunction for diatomic hydrogen and a linear four-hydrogen chain with a bond length of 0.75 $\AA$ described with an STO-3G basis using the OpenFermion~\cite{OpenFermion} and the OpenFermion-Psi4 plugin~\cite{parrish2017psi4}.  One hundred different corrupted RDMs were constructed by applying zero-mean Gaussian noise with variance $\epsilon^{2}$
\begin{align}
^{2}D_{rs}^{pq}{}_{\mathrm{measured}} = {}^{2}D_{rs}^{pq} + \mathcal{N}(0, \epsilon^{2}).
\end{align}
This error model is bias-free because the energy is linearly proportional to the $2$-RDM and has variance proportional to the error added to the $2$-RDM elements. For each of the one hundred corrupted density matrices we solve for a projected $2$-RDM with the positive projection, positive projection with fixed-trace, SDP $n$-representability reconstruction, and iterative $n$-representability projection techniques.  For each method we find the mean-square-error (MSE) of the aforementioned observables over the projected $2$-RDMs as a function of the noise parameter $\epsilon$.

\begin{figure}
    \centering
    \includegraphics[width=8.5cm]{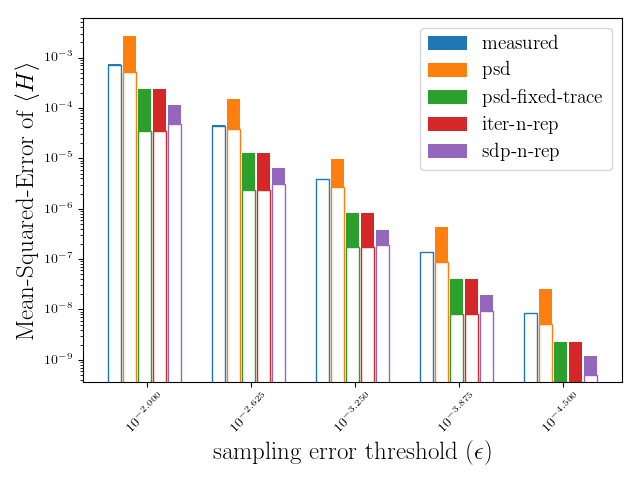}
    \includegraphics[width=8.5cm]{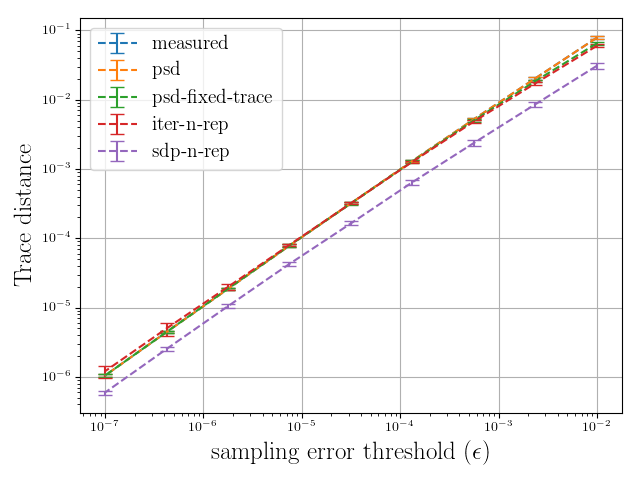}   
    \caption{left) the mean-squared-error (MSE) in the energy estimator for one-hundred samples.  MSE is decomposed into variance and bias squared in order to demonstrate how the projection techniques reduce the variance on the distribution of the estimators at the cost of inducing a bias.  The distribution of estimators without projection (labeled as measured) show no bias as expected based on the Gaussian error model and the fact the energy is a linear functional of the $2$-RDM. right) Trace distance of the $2$-RDM--measured or purified--from the true $2$-RDM.}
    \label{fig:RDMaccuracy}
\end{figure}

Figure~\ref{fig:RDMaccuracy} contains a plot of the MSE of the energy estimator for $\mathrm{H}_{2}$ decomposed into its variance and bias components, and a plot of the average trace distance of the reconstructed $2$-RDMs from the true $2$-RDM of $\mathrm{H}_{4}$.  The solid bars in the MSE plot are the squared bias component of the MSE while the transparent bars are the variance component.  In general, the $n$-representability inspired projection techniques decrease the variance of the energy estimator but introduce a bias.  Similar MSE plots are shown for $\langle S^{2} \rangle$, $\langle S_{z} \rangle$, and $\langle n \rangle$ in Appendix~\ref{appendix:constrained_observables}.  The expected value for $S_{z}$, $S^{2}$, and $n$ shows zero mean-squared-error for the SDP-based projection technique because these values are added as constraints to the semidefinite program.  We refer to the correction of the three aforementioned expected values as restoration of physicality--i.e. the particle number expectation is what is expected for an isolated system.  The reduced trace distance for the SDP projected $2$-RDMs for $\mathrm{H}_{4}$ indicates that the physicality constraints are important for removing errors from $2$-RDMs measured from a quantum resource.
\subsubsection{Reconstruction of marginals from the variational channel state model}
One appealing application of projection techniques based on $n$-representability is purification of states corrupted by an error channel. 
The SDP $n$-representability reconstruction procedure ensures physicality of the states by ensuring known symmetries are preserved by formulating the projection as a constrained optimization.  To verify this we used the SDP $n$-representability method in conjunction with the variational channel state error models presented in Reference~\cite{McClean2016} to demonstrate the restoration of physicality.   The variational channel state model is implemented by corrupting a pure state $\vert \psi \rangle$ with a channel described in Kraus operator form.  We consider uniform uncorrelated single-qubit error channels associated with dephasing, amplitude damping and dephasing, and depolarizing noise.  The dephasing and amplitude damping channels are parameterized with the assumption that 5\% of the coherence time has elapsed with respect to $T_{1}$ and $T_{2}$.  For the depolarizing channel, the Kraus operators are constructed assuming 5\% of the dephasing time $T_{2}$ has elapsed.  For each point along the binding curve of diatomic hydrogen the action of the channel on the pure-state is calculated as
\begin{align}
\rho_{\mathrm{channel}} = \sum_{i=1}^{M}K_{i} \rho_{\mathrm{pure}} K_{i}^{\dagger}.
\end{align}
Each ensemble $2$-RDM is then reconstructed with $2$-positive $n$-representability conditions using the SDP projection technique where the error matrices are set as the spin-adapted components of the $2$-RDM associated with $\rho_{\mathrm{channel}}$.  Spin adapting eliminates the need to explicitly enforce the antisymmetry of the $2$-RDM elements and thus reduces the total number of constraints in the SDP.

\begin{figure}
    \includegraphics[width=8.5cm]{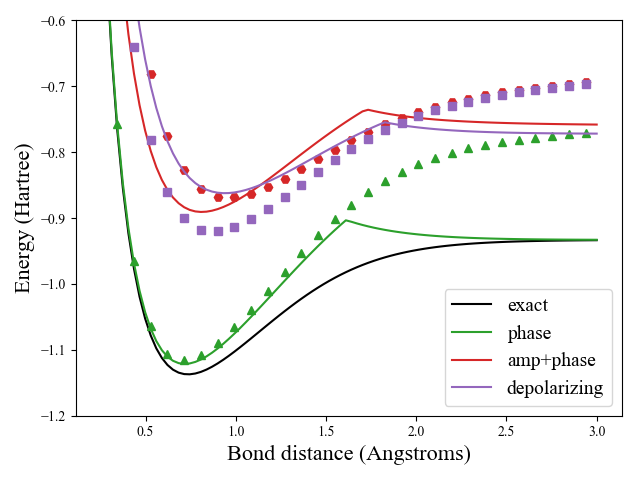}
    \caption{Energy curves for molecular hydrogen after action of three error channels.  The uniform uncorrelated single qubit error channels applied in the variational channel state model assume the entire circuit is executed within 5\% of the total coherence time.  The solid lines are the curves without $n$-representability reconstruction while the markers of the same color indicate the reconstruction under exact $n$-representability conditions. The black curve, depicted by $\textit{exact}$ is the true ground state energy curve.  The label \textit{amp} refers to the single-qubit amplitude damping channel associated with $T_{1}$ time, \textit{phase} refers to the single-qubit dephasing channel associated with $T_{2}$ time, and \textit{depolarizing} is the associated with the single-qubit depolarizing channel.\label{fig:nrep_binding_channels}} 
\end{figure}

The energy of the the $\mathrm{H}_{2}$ system under the action of each separate channel is plotted in Figure~\ref{fig:nrep_binding_channels} along with the energy computed as a functional of the reconstructed $2$-RDM.  The kinks in the dephasing and amplitude $+$ dephasing curves are associated with a spin-symmetry breaking, where the channels produced a mixed state dominated by a triplet state.  The discontinuity in the depolarizing channel curve is associated with the channel state switching to be a mixed state with a large component of singlet character.  The markers without a line in Figure~\ref{fig:nrep_binding_channels} are the reconstructed $2$-RDM with $\langle S_{z} \rangle = 0$ and $\langle S^{2} \rangle = 0$ imposed by linear constraints on the $2$-RDM associated with $S$-representability~\cite{PhysRevA.72.052505}.  Naturally the binding curves are now smooth as a function of bond distance.  Though physicality is recovered by projecting onto the closest marginal with fixed symmetries, the energy increases at distances greater than 1.5 Angstroms for each error model and the potential energy minimum is shifted by -0.03 $\AA$ for the dephasing channel, 0.062 $\AA$ for the dephasing and relaxation channel, and 0.186 $\AA$ for the depolarizing channel.  The energy increase is due to constraining the projected 2-RDM to have the correct spin-symmetry when the the error channel, applied through the variational channel state model, has switched the ground state symmetry from a singlet to a triplet~\cite{McClean2016}.  The increase of energy upon restoration of a symmetry is a well known effect in chemical systems and condensed matter systems.  However, the qualitative improvement in the nuclear potential energy surface and the implications for forces derived from such a surface are more important than the energy increase that is incurred.  The relative error between corrected 2-RDM and the uncorrected 2-RDM at the maximum bond distance considered in this work (3.0 $\AA$) is 18 percent for the dephasing channel, 8.7 percent for the dephasing and relaxation channel, and 9.9 percent for the depolarizing channel.

\section{Use cases of RDMs in augmenting energy expectation through perturbation theory}\label{sec:rdm_pertubation}
In this section, we briefly mention some of the additional use cases within chemistry for correcting the reduced density matrices beyond simple estimation of the energy, and review techniques that allow one to utilize them with only the $2$-RDM.  In traditional approaches to electronic structure on classical computers, the solution of the electronic structure problem within an active space often lacks the contributions from so-called dynamical correlation.  This part of the correlation stems largely from the electronic cusp contributions and represents a low-rank interaction in many high-lying orbitals.  Multi-reference perturbation theory techniques have been found to offer a good balance between cost and accuracy for including these contributions, and many approaches have been developed in this regard including complete active space second order perturbation theory (CASPT2)~\cite{Andersson_1990,Andersson_1992,Celani_2000}, multi-reference Moller-Plesset theory~\cite{Hirao_1992}, n-electron valence perturbation theory (NEVPT2)~\cite{Angeli_2001,Angeli_2002,Angeli_2006}, canonical transformation theory (CT)~\cite{Yanai_2006,Yanai_2007,Neuscamman_2010}, perturbative explicitly correlated corrections ($[2]_{\text{R}12}$)~\cite{Torheyden_2009,Kong_2010}, and corrected multi-reference CI (MRCI+Q)~\cite{Werner_1982,Werner_1988}.

In understanding how these techniques may be utilized within the quantum domain, one is most interested in those that are compatible with the measurements one expects to be able to feasible make on a quantum computer.  This means that explicit knowledge of the determinant decomposition of a wavefunction cannot be required, and instead one prefers methods that require only the $k$-RDM of the electronic system, where hopefully $k$ is small.  Of those discussed, NEVPT2, $[2]_{\text{R}12}$, and CT can be applied in this way.

For example, in NEVPT2, the effective equations may be derived entirely with the use of the $4$-RDM, and it does not require knowledge of determinant decomposition or full electronic wavefunction.  However the $4$-RDM is a relatively expensive quantity to estimate, and thus cumulant based approximations to the $4$-RDM have been developed such that only the $2$-RDM is required to correct the energy perturbatively~\cite{Zgid:2009,Neuscamman_2010}.  

The cumulant expansions decompose the reduced density matrices into their non-separable (connected) components and separable unconnected components, and are quite useful for both developing approximations and enhancing understanding. A convenient notation for expressing these expansions is given by the Grassmann wedge product defined generally by
\begin{align}
 a \wedge b = \left( \frac{1}{N!} \right)^2 \sum_{\pi, \sigma} \epsilon(\pi) \epsilon(\sigma) \ \pi \  \sigma \ a \otimes b
\end{align}
where $\pi$ and $\sigma$ are permutations on the upper and lower indices of the tensor $a\otimes b$ and $\epsilon$ denotes the parity of each permutation.  As an example one might consider the wedge product of a cumulant matrix with itself
\begin{align}
 \left[{}^1\Delta \wedge {}^1 \Delta\right]^{i_1 i_2}_{j_1 j_2} = \frac{1}{2} \left( {}^1\Delta^{i_1}_{j_1} {}^1\Delta^{i_2}_{j_2} - {}^1\Delta^{i_1}_{j_2} {}^1\Delta^{i_2}_{j_1} \right).
\end{align}
With this notation, the reduced density matrices up to $k=4$ may be expressed in terms of the cumulant expansions:
\begin{align}
 {}^{1}D &= {}^{1}\Delta \\
 {}^{2}D &= {}^{2}\Delta + {}^{1}\Delta \wedge {}^{1}\Delta \\
 {}^{3}D &= {}^{3}\Delta + 3 {}^{2}\Delta \wedge {}^{1}\Delta + {}^{1}\Delta \wedge {}^{1}\Delta \wedge {}^{1}\Delta \\
 {}^{4}D &= {}^{4}\Delta + 4 {}^{3}\Delta \wedge {}^{1}\Delta + 3 {}^{2}\Delta \wedge {}^{2}\Delta \notag \\
 &+ 6 {}^{2}\Delta \wedge {}^{1}\Delta \wedge {}^{1}\Delta
  + {}^{1}\Delta \wedge {}^{1}\Delta \wedge {}^{1}\Delta \wedge {}^{1}\Delta.
\end{align}
These methods that neglect contributions from the 3- and 4-cumulant dramatically reduce the number of samples that would be required to estimate the energy, however the approximations introduce some error in the corrections.  An additional factor is the consideration of the impact of measurement noise on the RDM in these numerical procedures.  While the purification techniques suggested in this draft are expected to mitigate some problems in this regard, the effect on an iterative procedure may be dramatic.  For this reason, NEVPT2 and its approximations may be the preferred method for use with quantum computers.  The strongly contracted equations that explicitly define the corrections to the energy are given in Appendix A of Ref. ~\cite{Angeli_2002}, and cumulant reconstruction methods may be used directly to form approximations as dictated in Ref. ~\cite{Zgid:2009}.  The above equations can be used to derive cumulant based approximations for up to the 4-RDM from using only the 2-RDM by setting the irreducible three- and four-particle components $\Delta^3$ and $\Delta^4$ to $0$.  Alternatively, more sophisticated approximation schemes have been developed in the context of RDM theory within traditional quantum chemistry~\cite{Mazziotti_2003,Greenman_2008}

\section{Conclusion}
Reducing the number of experiments required in the partial tomography step of VQE and other hybrid algorithms is necessary for hybrid classical/quantum algorithms to become useful simulation tools.  In this work we proposed using representability conditions on fermionic marginals as a route towards reducing the number of measurements required in a VQE operator averaging step.   Directly measuring the marginals provides the information required to integrate the results from hybrid quantum algorithms with classical quantum simulation methods.  

Fermionic representability conditions were used in two majors ways: 1) re-expressing the Pauli sum Hamiltonian in a form where the total variance is minimized given a fixed state and 2) designing projection techniques based on necessary conditions on $2$-marginals.   Both techniques have shown significant promise toward minimizing the total number of measurements and reducing stochastic noise seen from sampling the quantum resources.  

The SDP projection techniques are especially attractive because they are constructed to return physical states.  We have observed the restoration of physicality when a pure-state is corrupted by single-qubit error channels.  More significant representability conditions based on positivity of elements of the $3$-RDM may provide more accurate reconstruction under noise.  Another area that is especially exciting is the application of pure state constraints in the reconstruction procedure.  The conditions discussed in this work do not constrain the marginal to be integrated from a pure state.  As a result, the positivity constraints likely do little for the preservation of physicality as compared to the equality representability constraints, such as fixed number operator and constrained total spin.  Pure-states could potentially further reduce systematic errors associated with gates.

We fully expect that further investigation of representability conditions with realistic systems, more realistic error models, and performant numerical implementations will demonstrate the utility of measuring marginals for quantum simulation using hybrid classical/quantum algorithms.
\section*{Acknowledgements}

The authors thank Marcus da Silva for helpful discussions about error models and estimating observables, Ding Nan and Nathan Wiebe for helpful conversations about techniques for reducing measurement variance, and Sergio Boixo for helpful conversations about concentration of measure in random states.

\bibliography{biblo,Mendeley}
\appendix
\section{Structure of RDMs}\label{appendix:rdms_struct}
The semidefinite program for reconstructing a noisy $2$-RDM has significant block-diagonal structure for chemical problems.  The block structure of the $2$-RDM reflects the symmetries of the Hamiltonian.  Therefore, the $2$-RDM can be blocked according to $\langle \hat{S}_{z} \rangle$, $\langle S^{2} \rangle$, $\langle n \rangle$, and any spatial symmetry groups.  For example, time-reversal symmetry implies spin-adapted $1$-RDM when a position space basis is used and provides additional constraints when momentum is a good quantum number.  For Gaussian basis sets commonly used in quantum chemistry the time-reversal invariant spin-free quantum chemical Hamiltonian implies the following block structure: the total $1$-RDM and $1$-Hole-RDM can be blocked into $\alpha$ and $\beta$ spin blocks
\begin{align}
^{1}D = \begin{pmatrix}
^{1}D_{\alpha} & 0 \\
0 & ^{1}D_{\beta}
\end{pmatrix}
\end{align}
and
\begin{align}
^{1}Q = \begin{pmatrix}
^{1}Q_{\alpha} & 0 \\
0 & ^{1}Q_{\beta}
\end{pmatrix}
\end{align}
while the two particle matrices can be blocked as follows
\begin{align}
^{2}D = \begin{pmatrix}
^{2}D_{\alpha, \alpha}^{\alpha, \alpha} & 0 & 0 & 0\\
0 & ^{2}D_{\beta, \beta}^{\beta, \beta} & 0 & 0 \\
0 & 0 & ^{2}D_{\alpha, \beta}^{\alpha, \beta} & 0 \\
0 & 0 & 0 & ^{2}D_{\beta, \alpha}^{\beta, \alpha}
\end{pmatrix}
\end{align}
where the blocks $^{2}D_{\alpha, \alpha}^{\alpha, \alpha}$, $^{2}D_{\beta, \beta}^{\beta, \beta}$, $^{2}D_{\alpha, \beta}^{\alpha, \beta}$ and  $^{2}D_{\beta, \alpha}^{\beta, \alpha}$ have linear sizes of $r_{s}^{2}$. In practice we can reduce the size of the $^{2}D_{\alpha, \alpha}^{\alpha, \alpha}$ and  $^{2}D_{\beta, \beta}^{\beta, \beta}$ blocks by noting that these tensors are spanned by $(r_{s} \text{ choose } 2)$ basis antisymmetric functions instead of $r_{s}^{2}$ symmetric functions.  The $^{2}D_{\beta, \alpha}^{\beta, \alpha}$ block is removed as it can be mapped in a one-to-one fashion to the $^{2}D_{\alpha, \beta}^{\alpha, \beta}$ block rendering it redundant.  The corresponding blocks in the Hamiltonian correspond to the antisymmeterized integrals.  The $^{2}Q$ matrix has the same block structure.  The $^{2}G$ matrix has slightly less block structure
\begin{align}
^{2}G = \begin{pmatrix}
^{2}G_{\alpha, \alpha}^{\alpha, \alpha} & ^{2}G_{\alpha, \alpha}^{\beta, \beta} & 0 & 0 \\
^{2}G_{\beta, \beta}^{\alpha, \alpha} & ^{2}G_{\beta, \beta}^{\beta, \beta} & 0 & 0 \\
0 & 0 & ^{2}G_{\alpha, \beta}^{\alpha, \beta} & 0 \\
0 & 0 & 0 & ^{2}G_{\beta, \alpha}^{\beta, \alpha} 
\end{pmatrix}
\end{align}
where the block sizes are $2r_{s}^{2}$, $r_{s}^{2}$, and $r_{s}^{2}$ respectively. 
\section{Observables from the $2$-RDM and $1$-RDM}\label{sec:rdm_observables}
In this work we make use of the fact that a number of important observables are linear functionals of the $2$-RDM and $1$-RDM.  In this section we enumerate these relationships for clarity.

The energy can of a chemical Hamiltonian can be expressed as a linear function of the $1$-RDM and $2$-RDM.  Consider a general chemical Hamiltonian in second quantization
\begin{align}
H = \sum_{ij} h_{ij}a_{i}^{\dagger}a_{j} + \frac{1}{2}\sum_{pqrs} V_{pqrs}a_{p}^{\dagger}a_{q}^{\dagger}a_{s}a_{r}
\end{align}
where $h_{ij}$ and $V_{pqrs}$ are the one- and two-electron integral tensors.  When evaluating the expected value of the Hamiltonian $\langle H \rangle$ the dependence on the $1$- and $2$-RDM naturally emerges
\begin{align}
\langle H\rangle =& \sum_{ij} h_{ij}\langle a_{i}^{\dagger}a_{j} \rangle + \frac{1}{2}\sum_{pqrs} V_{pqrs}\langle a_{p}^{\dagger}a_{q}^{\dagger}a_{s}a_{r} \rangle \nonumber \\
=& \sum_{ij} h_{ij} {}^{1}D_{j}^{i} + \frac{1}{2}\sum_{pqrs} V_{pqrs} {}^{2}D_{rs}^{pq}.
\end{align}

For the calculations depicted in Figure~\ref{fig:nrep_binding_channels} the total angular momentum $S^{2}$, projected angular momentum $S_{z}$, and the particle number $n$ operators are used as linear constraints in the semidefinite program.  Just like the energy, these operators are linear functionals of the $1$- and $2$-RDMs.  To see this we express each component of the aforementioned operators as sums of fermionic operators resulting in polynomials of rank-$4$ and rank-$2$.
\begin{align}
n =& \sum_{i = 1}^{m} a_{i}^{\dagger}a_{i} \\
S_{z} =& \frac{1}{2} \sum_{i=1}^{m/2} \left(a_{i, \alpha}^{\dagger}a_{i, \alpha} -a_{i, \beta}^{\dagger}a_{i, \beta} \right)\\
S^{2} =& S^{-}S^{+} +  S_{z}^{2}  + S_{z}  
\end{align}
where 
\begin{align}
S^{-} =& \sum_{i=1}^{m / 2} a_{i, \beta}^{\dagger}a_{i, \alpha} \\
S^{+} =& \sum_{i=1}^{m / 2} a_{i, \alpha}^{\dagger}a_{i, \beta},
\end{align}
$m$ is the total number of spin orbitals, and $\alpha$ ($\beta$) denotes the two eigenfunctions of the $z$-angular momentum operator for a single fermion.  The expected value of each operator can be determined by summing over the indicated elements of the $2$-RDM and $1$-RDM.
\section{Mapping Conditions and Trace Conditions}\label{appendix:mapping_conditions}
The linear constraints in the SDP-projection semidefinite program include a trace constraint on the $2$-RDM, mappings between the $1$-RDM, $1$-hole-RDM, $2$-RDM and $2$-hole-RDM, and the $2$-RDM to the $2$-particle-hole-RDM.  These mappings between the marginals can be derived by rearranging the fermionic ladder operators resulting in the following matrix element equalities:
\begin{align}
^{1}D^{p}_{q} \;+\; ^{1}Q_{p}^{q} = \delta_{p, q}
\end{align}
\begin{align}
^{2}D_{r,s}^{p,q} =& \;^{1}D_{r}^{p}\delta_{s}^{q} + ^{1}D_{s}^{q}\delta_{r}^{p} \nonumber \\
-& \left( \;^{1}D_{s}^{p}\delta_{r}^{q} + \;^{1}D_{r}^{q}\delta_{s}^{p} \right) \nonumber \\
-& \delta_{s}^{p}\delta_{r}^{q} + \delta_{r}^{p}\delta_{s}^{q} \nonumber \\
+& \;^{2}Q_{p, q}^{r, s}
\end{align}
\begin{align}
^{2}G_{r, s}^{p, q} = \delta_{s}^{q} \;^{1}D_{r}^{p} - \;^{2}D_{q, r}^{p, s}.
\end{align}
The contraction relation between the elements of the $2$-RDM and $1$-RDM
\begin{align}
\sum_{i}\;^{2}D^{p, i}_{r, i} = (n - 1) \; ^{1}D_{r}^{p}
\end{align}
is included in the set of linear constraints.
\section{Computational Implementation of the Reconstruction Problem}
The reconstruction problem outlined in Section \ref{sec:reconstruct} is formulated as a semidefinite program.  Unique to this program is the sparsity of each constraint relative to the total number of variables in the program.  A class of SDP solvers using the augmented Lagrangian technique have been shown to efficiently solve SDPs of this form in quantum chemistry and condensed matter~\cite{baumgratz2012lower, rubin2015strong, rubin2014comparison, burer2003nonlinear, burer2005local, povh2006boundary, mazziotti2011large}.
The primal semidefinite program is mathematically stated as 
\begin{align}\label{Primal}
    \mathrm{min}\;\; \langle C, X \rangle \\
    \mathrm{s.t} \;\; \langle A, X \rangle = b \; ; \; X\succeq 0
\end{align}
where $C$ is in the space of symmetric matrices ($C \in \mathcal{S}_{n}$), $X$ is in the space of positive semidefinite matrices ($X \in \mathcal{S}_{n}^{+}$), $\langle \cdot , \cdot \rangle$ is defined as the trace inner product $\mathrm{Tr}[C \cdot X]$, $b$ is a vector in $\mathcal{R}^{m}$, and $A$ is the matrix of constraints.  The conjugate dual of the primal
\begin{align}
    \mathrm{max}\;\; b^{T} y \\
    \mathrm{s.t} \;\; S = C - y^{T}A \; ; \; S\succeq 0
\end{align}
the matrix $X$ is the primal variable and the pair $(y,S)$ are the dual variables.  In this work we use the boundary point method to solve the augmented Lagrangian dual to the SDP~\cite{povh2006boundary}.  The total boundary point algorithm is as follows
\begin{outline}
    \item{Repeat until $\delta_{\mathrm{outer}}  < \epsilon_{\mathrm{outer}}$}
    \begin{outline}
        \item{Repeat until $\delta_{\mathrm{inner}}  < \epsilon_{\mathrm{inner}}$}
        \begin{outline}
            \item{solve for $y^{k}$: $A(A^{T}y) = A(Z^{k} + C + \frac{1}{\sigma}X^{k} ) - \frac{1}{\sigma}b$}
            \item{Positive projection step: $W = A^{T}y^{k} - C - \frac{1}{\sigma}X^{k}$; $Z^{k} = W_{+}$ ; $V^{k} = W_{-}$ }
            \item{$\delta_{\mathrm{inner}} = || \langle A, V^{k} \rangle - b||$}
        \end{outline}
    \end{outline}
    \item{$X^{k+1} = V^{k}$}
    \item{$k = k+1$; $\delta_{\mathrm{outer}} = ||S^{k} - A^{T}y^{k} + C||$}
    \item{update $\sigma$}
\end{outline}
Here the positive and negative projections $W_{+}$ and $W_{-}$ are determined by the minimization
\begin{align}
W_{+} = \mathrm{argmin}_{U\succeq 0} || W - U ||
\end{align}
which corresponds to generating the $W_{+}$ by an eigenvalue decomposition and selecting positive eigenvalues along with their associated eigenvectors to generate the positive projection
\begin{align}
W_{+} = \sum_{i}\lambda_{i}^{+}|\phi_{i}\rangle \langle \phi_{i}| \;\;\;W_{-} = \sum_{j}\lambda_{j}^{-}|\phi_{j}\rangle \langle \phi_{j}|
\end{align}  
The computationally expensive task is the determination of $y$ in the inner minimization problem.  As $AA^{T}$ does not change its Cholesky decomposition, it can be formed prior to the calculation and then used to back-solve for $y^{k}$.  Using the backsolve method for the inner loop requires only one step.  For larger problems, we can solve the inner loop with the conjugate gradient method and thus must set the inner stopping $\epsilon_{\mathrm{inner}}$ condition significantly below the outer stopping condition $\epsilon_{\mathrm{outer}}$.  For all SDPs, we use the $L2$-norm of the primal error $||\langle A | X \rangle - b||_{2}$ as $\delta_{\mathrm{outer}}$.  All SDPs were stopped when $\delta_{\mathrm{outer}}$ fell below $1.0E-8$ or the number of outer iterations reached five-thousand.
\section{Constrained Observables}\label{appendix:constrained_observables}
To further examine the effects of the four projection techniques proposed we examined the mean-squared-error as a function of noise in the Gaussian error model and type of projection procedure used.  The Gaussian error model does not preclude a positive semidefinite $2$-RDM and thus restoration of physical symmetries such as positive-semidefinite-ness, constrained spin-, and particle-numbers are expected to increase the observed energy of the $2$-RDM with respect to the uncorrected noisy $2$-RDM.  
\begin{figure}
    \centering
    \includegraphics[width=8.5cm]{mse_var_bias2_energy_h2.png}
    \includegraphics[width=8.5cm]{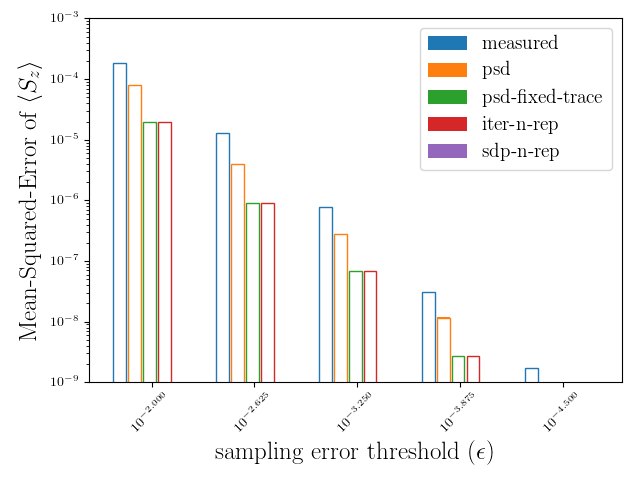}   
    \includegraphics[width=8.5cm]{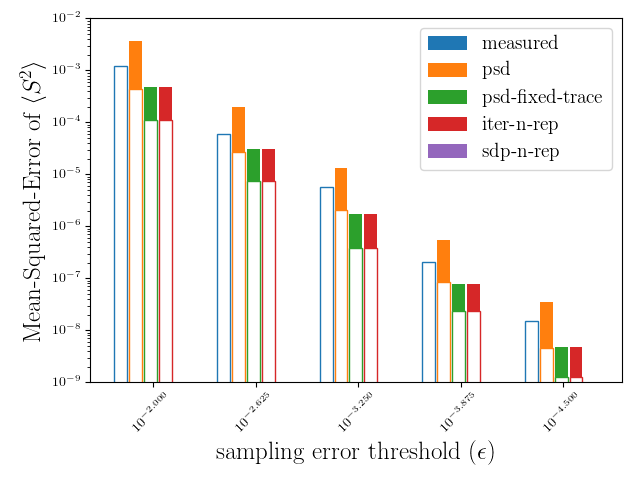}    
    \includegraphics[width=8.5cm]{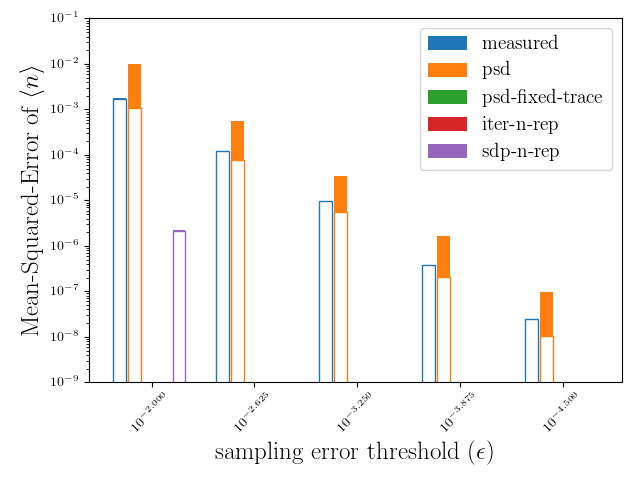}    
    \caption{The mean-squared-error (MSE) in the estimators for energy $\langle H \rangle$, total spin $\langle S^{2} \rangle$, projected spin $\langle S_{z} \rangle$, and particle number $\langle n \rangle$ for $\mathrm{H}_{2}$ over one-hundred samples.  MSE is decomposed into variance (clear bars) and bias (solid bars). }
    \label{fig:mse_h2}
\end{figure}
\begin{figure}
    \centering
    \includegraphics[width=8.5cm]{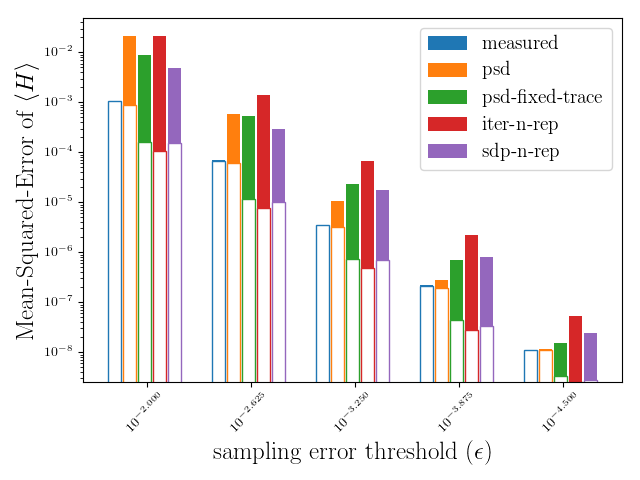}
    \includegraphics[width=8.5cm]{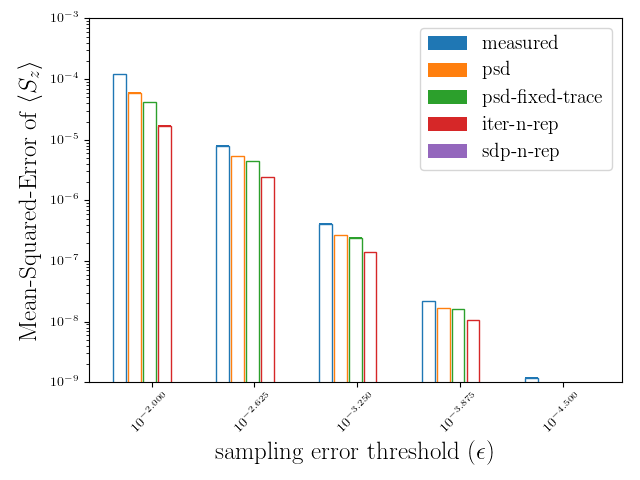}   
    \includegraphics[width=8.5cm]{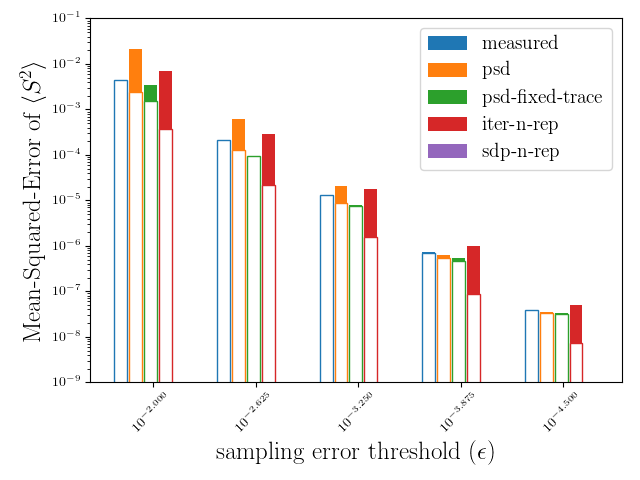}    
    \includegraphics[width=8.5cm]{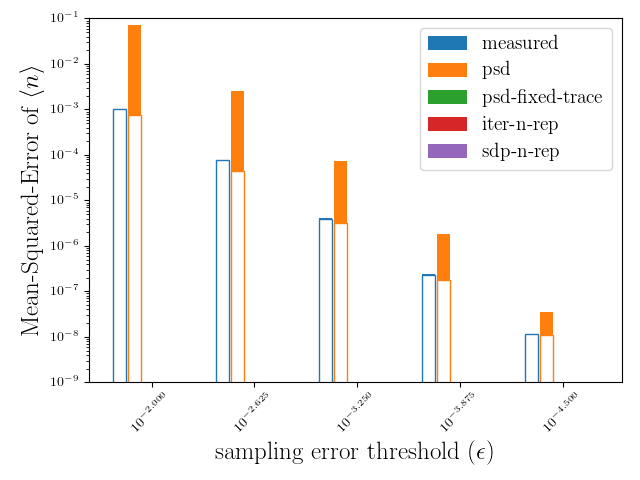}    
    \caption{The mean-squared-error (MSE) in the estimators for energy $\langle H \rangle$, total spin $\langle S^{2} \rangle$, projected spin $\langle S_{z} \rangle$, and particle number $\langle n \rangle$ for $\mathrm{H}_{4}$ over one-hundred samples.  MSE is decomposed into variance (clear bars) and bias (solid bars).}
    \label{fig:mse_h2}
\end{figure}
\end{document}